\documentclass[preprint,showpacs,showkeys,preprintnumbers,amsmath,amssymb,nofootinbib,aps]{revtex4}
\usepackage{amssymb}
\usepackage{amsmath}
\usepackage{graphicx}
\usepackage{dcolumn}
\usepackage{multirow}
\usepackage{bm}
\usepackage{epsfig}
\usepackage[caption=false]{subfig}
\usepackage[all]{xy}
\usepackage[bookmarks=false]{hyperref}
\xyoption{import,color,arc}

\newcommand\beq{\begin{eqnarray}}
\newcommand\eeq{\end{eqnarray}} 
 
\newcommand\eq[1]{Eq. (\ref{eq:#1})} 
\newcommand\nn{\nonumber}
 
\newcommand\Eq[1]{Eq.~\ref{eq:#1}}
\newcommand\Fig[1]{Fig.~\ref{fig:#1}}
\newcommand\Sec[1]{Sec.~\ref{sec:#1}}
\newcommand\Appendix[1]{Appendix~\ref{sec:#1}}

\newcommand\Tab[1]{Table~\ref{tab:#1}}

\newcommand\calA{\mathcal A}
\newcommand\calB{\mathcal B}

\newcommand\calF{\mathcal F}

\newcommand\calO{\mathcal O}
\newcommand\calT{\mathcal T}

\newcommand\calI{\mathcal I}
\newcommand\calH{\mathcal H}

\newcommand\calQ{\mathcal Q}
\newcommand\calS{\mathcal S}
\newcommand\calM{\mathcal M}
\newcommand\calW{\mathcal W}

\newcommand\tlam{{\lambda^{\textrm{PQ}}_k}}
\newcommand\teye{\textrm{eye}}
\newcommand\tsea{\textrm{sea}}
\makeatletter

\newcommand{\Rmnum}[1]{\expandafter\@slowromancap\romannumeral #1@}
\makeatother

\begin{document}

\title{Matrix elements of $\Delta B=0$ operators in heavy hadron chiral perturbation theory} 

\begin{abstract}
We study the light-quark mass and spatial volume dependence of the matrix elements of 
$\Delta B=0$ four-quark operators relevant for 
the determination of $V_{ub}$ and the lifetime ratios of single-$b$ hadrons. 
To this end, one-loop diagrams are computed 
in the framework of heavy hadron chiral perturbation theory with  partially 
quenched formalism for three light-quark flavors in the isospin limit; 
flavor-connected and -disconnected diagrams are carefully analyzed. 
These calculations include the leading light-quark flavor and heavy-quark spin symmetry breaking effects 
in the heavy hadron spectrum. 
Our results can be used in the chiral extrapolation of lattice calculations of 
the matrix elements to the physical light-quark masses and to infinite volume. 
To provide insight on such chiral extrapolation, 
we evaluate the one-loop contributions to the matrix elements containing external $B_d$, $B_s$ mesons 
and $\Lambda_b$ baryon in the QCD limit, 
where sea and valence quark masses become equal. 
In particular, we find that the matrix elements of the $\lambda_3$ flavor-octet operators 
with an external $B_d$ meson receive the contributions solely from connected diagrams 
in which current lattice techniques are capable of precise determination of the matrix elements. 
Finite volume effects are at most a few percent for typical lattice sizes and pion masses. 
\end{abstract}

\author{Jong-Wan Lee}
\email{jwlee2@ccny.cuny.edu}
\affiliation{Department of Physics, The City College of New York, New York, New York 10031, USA}





\date{\today}

\maketitle

\section{Introduction}
\label{sec:introduction}

In inclusive decays of single-$b$ hadrons, the effects of spectator quarks 
(light constituent quarks in the hadrons) 
play an important role in extracting their decay widths and the lifetimes of different species 
as those quarks participate in the weak process. 
In particular, 
for a number of years there have been suggestions that the lifetime of the $\Lambda_b$ baryon 
had significant corrections 
from the spectator effects \cite{Neubert:1996we,Gabbiani:2003pq}, 
which might explain the small experimental value of the lifetime ratio $\tau(\Lambda_b)/\tau(B_d)$. 
Recent measurements tend to favor more natural values \cite{Amhis:2012bh} :
\beq
\frac{\tau(B^+)}{\tau(B_d)}=1.079\pm 0.007, ~~~~~
\frac{\tau(B_s)}{\tau(B_d)}=0.993\pm 0.009, ~~~~~
\frac{\tau(\Lambda_b)}{\tau(B_d)}=0.930\pm 0.020.
\eeq
Theoretical determinations of these lifetime ratios rely on 
the heavy-quark expansion (HQE) for inclusive decays with one heavy quark 
in the final state \cite{Chay:1990da,Bigi:1992su,Bigi:1993fe,Manohar:1993qn}. 
Although the $\Lambda_b$ lifetime is no longer a puzzle, 
it is desired to improve the theoretical determinations of the lifetime ratios 
since their uncertainties are larger than those of experimental measurements 
by an order of magnitude except $\tau(B_s)/\tau(B_d)$ \cite{Stone:2014pra}. 
In particular, the dominant source of systematic uncertainties in the HQE predictions 
are the size of the nonperturbative matrix elements of the dimension-six 
$\Delta B=0$\footnote{$\Delta B$ denotes the change of $b$-quark number by the operators.
} 
four-quark operators containing spectator quarks \cite{Lenz:2014jha}, 
which are poorly known from exploratory lattice studies \cite{DiPierro:1998ty,Becirevic:2001fy,
DiPierro:1999tb} 
and QCD sum rules \cite{Baek:1998vk,Cheng:1998ia,Huang:1999xj,Tarantino:2007nf}. 
Thus, the determination of these matrix elements from the state-of-the-art 
lattice calculations will significantly reduce 
the uncertainties and play an essential role in the precision test of the HQE.\footnote{
As emphasized in a recent review paper about the current status of the HQE \cite{Lenz:2014jha}, 
the HQE is now in the era of precision tests as 
the significant discrepancies between the HQE predictions and experimental measurements 
are resolved for the $\Lambda_b$ lifetime and 
the semileptonic branching ratios of $B$ mesons. 
Moreover, the validity of the HQE is strongly supported from the fact that 
the recent measurement of the decay rate differences $\Delta \Gamma_s$ \cite{Amhis:2012bh} 
perfectly agrees with the HQE predictions \cite{Lenz:2011ti}. 
}

The theoretical calculation of the decay width can be done systematically
using the optical theorem, which relates the total decay rate of the single-$b$ hadron $H_b$ 
to the imaginary part of the matrix element of the forward scattering amplitude, 
\beq
\Gamma(H_b\rightarrow X)=\frac{1}{M_{H_b}}\textrm{Im}\langle H_b|{\bf T}|H_b\rangle, 
\eeq
where $M_{H_b}$ is the hadron mass. Here the transition operator ${\bf T}$ is given by
\beq
{\bf T}=i\int d^4xT\lbrace \mathcal{L}_{eff}(x),\mathcal{L}_{eff}(0)\rbrace,
\eeq
where $\mathcal{L}_{eff}$ represents the effective $\Delta B=1$ weak Lagrangian 
renormalized at the scale of $b$-quark mass, $\mu=m_b$. 
Since the energy release is large ($\sim m_b$) in $b$-hadron decays, 
the operator product expansion (OPE) is applicable to construct local 
$\Delta B=0$ operators having the same quantum numbers, 
with suppression by inverse powers of $m_b$ accompanying increasing operator 
dimensions.\footnote{
It is well known that the OPE breaks down when one calculates 
differential inclusive $B$-meson decay distributions near the endpoint region in phase space, 
where a proper treatment of forward matrix elements of 
nonlocal operators is required \cite{Neubert:1993ch,Neubert:1993um,Bigi:1993ex}. 
Although the nonlocal operators typically reduce to local ones 
if the decay distributions are integrated over all phase space, 
there are cases such as the inclusive radiative decay $B\to X_s\gamma$ 
in which certain nonlocal matrix elements have sizeable corrections 
to the total decay rate \cite{Lee:2006wn,Benzke:2010js}. 
However, we note that in the case of the total inclusive decay rates of single-$b$ hadrons 
(i.e. all final states are summed), 
which is relevant for the determination of lifetimes, 
it is believed that the standard OPE is applicable. 
}
Using the OPE, one can write the decay width by
\beq
\Gamma(H_b\rightarrow X)=\frac{G_F^2 m_b^5}{192\pi^3}\frac{1}{2M_{H_b}}
\sum_k \frac{c_k(\mu)}{m_b^{k-3}}\langle H_b|\mathcal{O}_k^{\Delta B=0}(\mu)|H_b\rangle,
\label{eq:deltaB0_ops}
\eeq
where $c_k$ and $\mathcal{O}_k^{\Delta B=0}$ represent Wilson coefficients 
containing relevant CKM matrix elements and dimension-$k$ $\Delta B=0$ local operators, respectively.
The forward matrix elements of the local operators in \Eq{deltaB0_ops} are systematically 
expanded with inverse power of $m_b$ in the framework of 
heavy-quark effective theory (HQET) \cite{Grinstein:1990mj,Eichten:1989zv,Georgi:1990um}. 
This is the basic structure of the HQE. 
In these expansions the leading-order term is universal for all kinds of $b$ hadrons.
It is well known that deviations of the lifetime ratios first arise from the terms of order $1/m_b^2$ 
which include the nonperturbative corrections to $\langle H_b |\bar{b}b| H_b\rangle$ in the HQET and
the subleading terms in the OPE, $\bar{b}g_s i\sigma_{\mu\nu} G^{\mu\nu}b$, whose dimension is five 
\cite{Bigi:1992su,Bigi:1993fe,Manohar:1993qn}. 
The spectator effects contribute to the process at the order of $1/m_b^3$, 
where the corresponding dimension-six $\Delta B=0$ four-quark operators in a convenient basis are given by
\cite{Neubert:1996we}\footnote{
For $s_\ell=0$ single-$b$ baryons, only the first two four-quark operators 
are sufficient to compute the spectator effects in the heavy-quark limit, because $\calO_{1}$ ($\calO_{2}$) 
are related to $\calO_{3}$ ($\calO_{4}$) up to $1/m_b$ corrections via a Fierz transformation 
thanks to heavy-quark symmetry. 
}
\beq
\mathcal{O}_{1,a}&=&\bar{b}^\alpha\gamma_\mu (1-\gamma_5)q^{a,\alpha}\bar{q}_a^\beta\gamma^\mu(1-\gamma_5)b^\beta, \nn \\
\mathcal{O}_{2,a}&=&\bar{b}^\alpha(1-\gamma_5)q^{a,\alpha}\bar{q}_a^\beta(1+\gamma_5)b^\beta, \nn \\
\mathcal{O}_{3,a}&=&\bar{b}^\alpha\gamma_\mu (1-\gamma_5)q^{a,\beta}\bar{q}_a^\alpha\gamma^\mu(1-\gamma_5)b^\beta, \nn\\
\mathcal{O}_{4,a}&=&\bar{b}^\alpha(1-\gamma_5)q^{a,\beta}\bar{q}_a^\alpha(1+\gamma_5)b^\beta,
\label{eq:four_quark_ops}
\eeq
where $b$ denotes a $b$-quark field and $q^a$ denotes a light-quark field of flavor $a$. 
The color indices $\alpha$, $\beta$ and the Lorentz indices $\mu$ are summed, 
but the flavor indices $a$ are not. 
In fact, although these effects are suppressed by an additional power of $1/m_b$,
the contributions are numerically enhanced by a phase-space factor of $16\pi^2$ \cite{Neubert:1996we}. 

It was also pointed out that the contribution of the four-quark operators in \Eq{four_quark_ops} 
can be significant 
to the inclusive decay $B\rightarrow X_u \ell \nu_\ell$
which is closely related to the determination of $V_{ub}$ \cite{Voloshin:2001xi}. 
The four-quark operators relevant to this semileptonic decay are
given by $\mathcal{O}_{1,u}-\mathcal{O}_{2,u}$ at the scale $\mu=m_b$ 
and their matrix elements can be parametrized by
\beq
\frac{1}{M_B}(\langle B|\mathcal{O}_{1,u}|B\rangle-\langle B|\mathcal{O}_{2,u}|B\rangle)
=f_B^2M_B(B_1-B_2),
\label{eq:me1}
\eeq
where $f_B$ and $M_B$ are the decay constant and the mass of $B$ meson, respectively, and $B_1$ and $B_2$ 
are phenomenological parameters called bag constants.
In the vacuum insertion approximation (or factorization) \cite{Shifman:1978bx,Shifman:1978by}, 
we have $B_1=B_2$ and the spectator effects in \Eq{me1} vanish.
However, the actual values of these matrix elements should be determined from nonperturbative calculations 
and can be much different than that from the factorization approximation. 
In Ref. \cite{Voloshin:2001xi}, for instance, the author showed that
the correction to the branching ratio, $\delta B(B\rightarrow X_u\ell \nu_\ell)$, by spectator effects 
substantially enhances the estimate of the corresponding errors in the hybrid expansion, 
e.g. a violation of the relation $B_1=B_2$ by $10\%$
makes the correction of the branching ratio twice larger than the estimated uncertainty 
in Ref. \cite{Neubert:2001ib}.

In \Eq{deltaB0_ops}, the Wilson coefficients $c_k(\mu)$ 
are determined from perturbative calculations 
while the matrix elements of local four-quark operators can be computed nonpertubatively, 
potentially from lattice QCD (see Ref. \cite{Lenz:2014jha} and references therein). 
From the point of view of lattice QCD, 
difficulties arise from the fact that the light quark $q$ in the operators can be 
different from the light valence-quark $q'$ in the single-$b$ hadrons, 
i.e. $q$'s are contracted to themselves, so called eye contractions \cite{Voloshin:2001xi}: 
the operators could mix with lower dimensional ones 
requiring a power-law subtraction. 
In addition, lattice calculations of the matrix elements including disconnected diagrams 
generally suffer from a noise problem. 
Although it is inevitable to avoid such difficulties in the calculation of the semileptonic decay, 
it is expected that the contributions of the eye contractions are negligible 
in the calculation of the $B$-meson lifetime ratios due to the light-quark flavor symmetry. 
In particular, we find that the matrix elements of the $\lambda_3$ flavor-octet operators 
involving an external $B_d$ meson, 
which are relevant to the determination of $\tau(B^+)/\tau(B_d)$, 
do not only exclude the eye contractions, 
but also the disconnected contributions\footnote{
More precisely, ``disconnected" means the flavor disconnected in which 
the light-flavor quarks in the operators are not connected with those in 
the external hadrons. 
Throughout this paper, similarly, 
the ``singlet" and ``octet" mean the flavor singlet and flavor octet, respectively.} 
at the next-to-leading order in chiral expansion. 

The purpose of this paper is to compute the one-loop chiral corrections, which are
functions of light-quark mass (or pion mass), to the matrix elements of 
$\Delta B=0$ four-quark operators in \Eq{four_quark_ops}. 
At present, the light-quark masses in many lattice studies are not physical due to limited computing resources, 
and thus the chiral extrapolation to the physical quark masses is essential 
to obtain the matrix elements with high precision. 
To do this, we consider heavy hadron chiral perturbation theory (HH$\chi$PT) 
\cite{Burdman:1992gh,Wise:1992hn,Yan:1992gz,Cho:1992gg,Cho:1992cf} 
at finite volume \cite{Arndt:2004bg,Detmold:2006gh,Detmold:2011rb} 
as the machinery to write the four-quark operators in terms of low-energy degrees of freedom 
such as heavy hadrons and Goldstone mesons.
Moreover, our calculation is performed in partially quenched chiral perturbation theory 
(PQ$\chi$PT) \cite{Bernard:1993sv} 
which is appropriate for current and foreseeable lattice calculations, 
as well as it naturally distinguishes disconnected diagrams from connected ones. 
Although the heavy quarks are assumed to be static, we consider the leading effect of 
various mass differences among $B$ mesons and single-$b$ baryons.
For $B$ mesons, the chiral corrections to the matrix elements of infinite volume QCD 
were previously calculated in Ref. \cite{Becirevic:2008us}.

The organization of this paper is as follows. 
In \Sec{pqhhcpt}, we briefly review heavy hadron chiral perturbation theory 
involving hadrons containing a single-$b$ quark in SU$(6|3)$ partially quenched theories.  
Mass differences in the heavy hadron spectrum, 
which originate from the light-quark flavor and heavy-quark spin 
symmetry breakings, are discussed in detail, 
where the relevant mass parameters enter the one-loop chiral computations. 
Section \ref{sec:four_quark_ops} contains the determination of $\Delta B=0$ four-quark operators 
in partially quenched heavy hadron chiral perturbation theory. 
In \Sec{matrix_elements}, we present our main results: 
calculation of the matrix elements of the four-quark operators 
including one-loop chiral contributions at infinite volume and in a finite spatial box, 
where connected and disconnected contributions are clearly separated. 
Integrals and sums appearing in the loop calculations 
are summarized in \Appendix{integrals_and_sums}, 
while coefficients for these calculations involving $B$ mesons and 
single-$b$ baryons are summarized in Appendixes \ref{sec:coefficients_B_meson} 
and \ref{sec:coefficients_B_baryon}, respectively. 
Moreover, the pion-mass dependence of chiral corrections and 
finite volume effects are discussed for the exemplified cases of 
external $B_d$, $B_s$ mesons and $\Lambda_b$ baryon. 
Finally, we conclude our work in \Sec{conclusion}.

\section{Partially Quenched Heavy Hadron Chiral Perturbation Theory}
\label{sec:pqhhcpt}

The interactions of heavy hadrons containing a heavy quark with Goldstone mesons 
can be described in the framework of chiral perturbation theory ($\chi$PT) combined 
with the HQET. 
The inclusion of the heavy-light mesons into $\chi$PT was first carried out in \cite{Burdman:1992gh,Wise:1992hn,Yan:1992gz} 
and extended to quenched and partially quenched theories in \cite{Sharpe:1995qp,Savage:2001jw}. 
The $1/M_P$ and chiral corrections were investigated in \cite{Boyd:1994pa}, 
where $M_P$ is the mass of the heavy-light pseudoscalar meson. 
The superfield appearing in this effective theory is \cite{Grinstein:1992qt} 
\beq
H^{(\bar{b})}_i&=&\left(P^{*(\bar{b})}_{i,\mu}\gamma^\mu-P^{(\bar{b})}_i\gamma_5\right)
\frac{1-\displaystyle{\not}v}{2},
\label{eq:field_H}
\eeq
where $P^{(\bar{b})}_i$ and $P^{*(\bar{b})}_{i,\mu}$ annihilate pseudoscalar and 
vector mesons containing an anti-$b$ quark and a light quark of flavor $i$. 
For convenience, the space-time variables in the fields and transformations do not 
explicitly appear throughout this paper (e.g. $H^{(\bar{b})}_i(x)=H^{(\bar{b})}_i$). 
In the heavy quark formalism, the momentum of such mesons is given as 
$p^\mu=M_P v^\mu+k^\mu$ with $k^\mu\ll M_P$, where $v^\mu$ is 
the $4$-velocity of the meson fields. 
Under a heavy-quark spin SU$(2)$ transformation $S_h$ and 
a light-flavor transformation $U$ 
[i.e., $U\in \textrm{SU}(3)$ for full QCD and $U\in \textrm{SU}(6|3)$ for partially quenched QCD (PQQCD)], 
the field $H^{(\bar{b})}$ transforms as
\beq
H^{(\bar{b})}_i \rightarrow U^{~j}_i H^{(\bar{b})}_j S_h^{\dagger}\nn.
\eeq
The conjugate field, which creates mesons containing an anti-$b$ quark and a light quark 
of flavor $i$, is defined as
\beq
\bar{H}_i^{(\bar{b})}=\gamma^0 H^{(\bar{b})\dagger}_i \gamma_0
=\frac{1-\displaystyle{\not}v}{2}\left(
P^{*(\bar{b})\dagger}_{i,\mu}\gamma^\mu+P^{(\bar{b})\dagger}_i\gamma_5
\right),
\eeq
and transforms under $S_h$ and $U$ as
\beq
\bar{H}_i^{(\bar{b})} \rightarrow S_h \bar{H}_j^{(\bar{b})} U^{\dagger j}_{~~i}.
\eeq

The inclusion of baryons containing a $b$-quark and two light quarks into $\chi$PT 
was first proposed in \cite{Yan:1992gz,Cho:1992gg, Cho:1992cf}, and the effective theory was generalized to 
the quenched and partially quenched theories in \cite{Chiladze:1997uq,Chen:2001yi,Tiburzi:2004kd}. 
In the limit $m_b\rightarrow\infty$, the heavy quark's spin decouples from the system and 
the total spin of the light degrees of freedom is conserved. 
Because of this property of the heavy-quark symmetry, one can classify baryons by 
the total spin quantum number of the light degrees of freedom, $s_\ell=0$ or $s_\ell=1$. 
These two types of baryons carrying light flavors $i$ and $j$ can be included into SU$(6|3)$ PQ$\chi$PT 
by introducing the corresponding interpolating fields at the quark level as 
\beq
\calT^\gamma_{ij}&\sim& b^{\gamma,c}\left(q^{\alpha,a}_i q^{\beta,b}_j 
+ q^{\beta,b}_i q^{\alpha,a}_j\right)\epsilon_{abc}(C\gamma_5)_{\alpha\beta}~\textrm{for}~ s_\ell=0,\nn \\
\calS^{\gamma,\mu}_{ij}&\sim& b^{\gamma,c}\left(q^{\alpha,a}_i q^{\beta,b}_j 
- q^{\beta,b}_i q^{\alpha,a}_j\right)\epsilon_{abc}(C\gamma^\mu)_{\alpha\beta}~\textrm{for}~ s_\ell=1,
\eeq 
where $C$ is the charge-conjugation matrix, $\alpha, \beta, \gamma$ are the Dirac indices, 
and $a,b,c$ are color indices. 
These flavor tensor interpolating fields satisfy
\beq
\label{eq:baryon_flavor}
\calT_{ij}=(-)^{\eta_i \eta_j} \calT_{ji},~~~\calS^{\mu}_{ij}
=(-1)^{1+\eta_i \eta_j} \calS^{\mu}_{ji},
\eeq
where the grading factor $\eta_k$ accounts for different statistics of quarks in PQQCD,
\beq
\eta_i=\left\{
\begin{array}{l}
1~\textrm{for}~k=1,2,3,4,5,6~\textrm{(valence and sea)},\\
0~\textrm{for}~k=7,8,9~\textrm{(ghost)}.\\
\end{array}
\right.
\eeq
The $\calT$ and $\calS$ fields form a ${\bf 39}$- and a ${\bf 42}$-dimensional representations of SU$(6|3)$, respectively. 
The baryon fields appearing in HH$\chi$PT 
have the same flavor properties of the corresponding interpolating fields as in \Eq{baryon_flavor}, 
where the combination of the heavy-quark spin $1/2$ and the total spin of light quarks $s_\ell$ leads us to the fact 
that $T$ baryon carries spin $1/2$, while $S$ baryon carries both spin $1/2$ and $3/2$ 
which are degenerate in the heavy-quark limit. 
If we restrict our attention to the pure valence-valence sector, 
we recover the familiar baryon tensors of QCD. 
For $s_\ell=0$, the baryon tensor $T^{(\textrm{valence-valence})}_{ij}$ is antisymmetric 
under the exchange of light-quark flavor indices 
and explicitly written as
\beq
T^{(\textrm{valence-valence})}_{ij}=\frac{1}{\sqrt{2}}\left(
\begin{array}{ccc}
0 & \Lambda_b & \Xi_b^{+1/2} \\
-\Lambda_b & 0 & \Xi_b^{-1/2} \\
-\Xi_b^{+1/2} & -\Xi_b^{-1/2} & 0 \\
\end{array}
\right),
\eeq
where the superscript indicates the $3$-component of the isospin. For $s_\ell=1$, the baryon 
tensor $S_{ij}$ is 
described by the superfield
\beq 
S^{ij}_\mu=\sqrt\frac{1}{3}(\gamma_\mu+v_\mu)\gamma^5 B^{ij}+B^{*ij}_\mu,
\label{eq:S_field}
\eeq
where $B_{ij}$ and $B_{ij}^{*\mu}$ are spin-$1/2$ and $3/2$ baryons, respectively.\footnote{
The baryon field $T_{ij}$ ($B_{ij}$) satisfies 
$\frac{1+\displaystyle{\not}v}{2}T_{ij}~(B_{ij})=T_{ij}~(B_{ij})$, 
while $B^{*\mu}_{ij}$ satisfies 
$\frac{1+\displaystyle{\not}v}{2}B^{*\mu}_{ij}=B^{*\mu}_{ij}$ and $\gamma^\mu B_\mu^*=0$.} 
In the valence-valence sector, the baryon tensor $B^{(\textrm{valence-valence})}_{ij}$ is 
symmetric under the exchange of light-quark flavor indices and explicitly written as 
\beq
B^{(\textrm{valence-valence})}_{ij}=\left(
\begin{array}{ccc}
\Sigma_b^{+1} & \frac{1}{\sqrt{2}}\Sigma_b^0 & \frac{1}{\sqrt{2}}\Xi_b^{'+1/2} \\
\frac{1}{\sqrt{2}}\Sigma_b^0 & \Sigma_b^{-1} & \frac{1}{\sqrt{2}}\Xi_b^{'-1/2} \\
\frac{1}{\sqrt{2}}\Xi_b^{'+1/2} & \frac{1}{\sqrt{2}}\Xi_b^{'-1/2} & \Omega_b \\
\end{array}
\right),
\eeq
and similarly for the $B_{ij}^{*\mu}$ fields. 
Under the heavy-quark spin transformation $S_h$ and the light-flavor transformation $U$, 
\beq
T_{ij}\longrightarrow S_h U_i^{~k} U_j^{~\ell} T_{k\ell},\nn \\
S_{ij}^\mu \longrightarrow S_h U_i^{~k} U_j^{~\ell} S^\mu_{k\ell}.
\eeq 
The conjugate fields, which create single-$b$ baryons, are denoted as $\bar{S}^\mu_{ij}$ and $\bar{T}_{ij}$. 

In PQ$\chi$PT, the leading-order (LO) chiral Lagrangian for the Goldstone mesons is
\beq
\mathcal{L}=\frac{f^2}{8}\textrm{str}[(\partial^\mu\Sigma^\dagger)(\partial_\mu\Sigma)+
2B_0(\Sigma^\dagger \calM_q+\calM_q^\dagger \Sigma)]+
\alpha_\Phi (\partial^\mu\Phi_0)(\partial_\mu\Phi_0)-M_0^2\Phi_0^2,
\label{eq:lagrangian_meson}
\eeq
where $\Sigma=\textrm{exp}(2i\Phi/f)=\xi^2$ is the nonlinear Goldstone field, 
with $\Phi$ being the matrix containing the standard Goldstone fields in the quark-flavor basis. 
In this work, we follow the supersymmetric formation of PQ$\chi$PT 
where the flavor group is graded \cite{Bernard:1993sv}. 
Therefore $\Sigma$ transforms 
linearly under SU$(6|3)_\textrm{L}\bigotimes \textrm{SU}(6|3)_\textrm{R}$,
\beq
\Sigma\rightarrow U_\textrm{L}\Sigma U^\dagger_\textrm{R},
\eeq
where $U_{\textrm{L}(\textrm{R})}\in \textrm{SU}(6|3)_{\textrm{L}(\textrm{R})}$ is the left-handed 
(right-handed) light-flavor transformation. 
The operation str[ ] means supertrace over the graded light-flavor indices. 
The low-energy constant $B_0$ is related to the chiral condensate by
\beq
B_0=-\frac{\langle 0 |\bar{u}u+\bar{d}d|0\rangle }{f^2},
\eeq
and the quark-mass matrix in the isospin limit is 
\beq
\calM_q=\textrm{diag}(m_u,m_u,m_s,m_j,m_j,m_r,m_u,m_u,m_s).
\label{eq:quark_mass_matrix}
\eeq
We keep the strange quark mass different from that of the up and down quarks in the valence, sea, and ghost sectors. 
Analogous to QCD, the strong U$(1)_A$ anomaly can give rise to the large mass of the singlet field 
$\Phi_0=\textrm{str}(\Phi)/\sqrt{6}$, 
i.e. the same size of the chiral symmetry breaking scale, and thus $\Phi_0$ can be integrated out, 
resulting in residual hairpin structures in the two-point correlation function of neutral mesons 
\cite{Sharpe:2001fh,Sharpe:2000bc}.

The Goldstone mesons couple to the above $B$ meson and single-$b$ baryon fields via the field $\xi$, 
which transforms as
\beq
\xi\rightarrow U_\textrm{L} \xi U^\dagger = U \xi U_\textrm{R}^\dagger,
\eeq
where $U$ is a function of $U_\textrm{L}$, $U_\textrm{R}$, and $\Phi$. 
The field $\xi$ can be used to construct the vector and axial-vector fields of meson,
\beq
V^\mu=\frac{i}{2}[\xi^\dagger \partial^\mu \xi + \xi \partial^\mu \xi^\dagger],~
A^\mu&=&\frac{i}{2}[\xi^\dagger \partial^\mu \xi - \xi \partial^\mu \xi^\dagger].
\eeq
As the vector field plays a role similar to a gauge field, 
the chiral covariant derivatives which act on the $B$ meson and single-$b$ baryon fields 
can be defined as
\beq
&&\mathcal{D}^\mu H_i^{(\bar{b})}=\partial^\mu H_i^{(\bar{b})} -i (V^\mu)^{~j}_i H_j^{(\bar{b})}, \nn \\
&&\mathcal{D}^\mu T_{ij} = \partial^\mu T_{ij} - i (V^\mu)^{~k}_i 
T_{kj} - i (-1)^{\eta_i(\eta_j+\eta_k)} (V^\mu)_j^{~k} T_{ik},\nn \\
&&\mathcal{D}^\mu S_{ij}^\nu=\partial^\mu S_{ij}^\nu -i (V^\mu)_i^{~k} 
S_{kj}^\nu -i (-1)^{\eta_i(\eta_j+\eta_k)} (V^\mu)_j^{~k} S_{ik}^\nu, \nn 
\eeq
The LO effective Lagrangian in the chiral and $1/M_B$ expansions is
\beq
\mathcal{L}_\textrm{HH$\chi$PT}^{(\textrm{LO})}&=&
-i\textrm{tr}_D[\bar{H}^{(\bar{b})i} v\cdot \mathcal{D} H^{(\bar{b})}_i]
-i(\bar{S}^\mu v\cdot \mathcal{D} S_\mu)+i(\bar{T} v\cdot \mathcal{D} T)
+\Delta^{(B)} (\bar{S}^\mu S_\mu) \\
&&+g_1\textrm{tr}_D[H^{(\bar{b})}_i \bar{H}^{(\bar{b})j} \gamma_\mu \gamma_5 A^{\mu,i}_j ]
+ig_2\epsilon_{\mu\nu\sigma\rho}(\bar{S}^\mu v^\nu A^{\sigma} S^{\rho})
+\sqrt{2}g_3\left[(\bar{T} A^{\mu} S_\mu) + (\bar{S}^\mu A_\mu T)\right], \nn
\label{eq:lagrangian_hhcpt}
\eeq
where $\textrm{tr}_D$ is the trace over Dirac space 
and $v_\mu$ is the velocity of the heavy hadrons. 
Note that the $B$-meson fields are of mass dimension $3/2$ like the single-$b$ baryon fields 
by absorbing $\sqrt{M_B}$ into $H^{(\bar{b})}_i$,
where $M_B$ is the mass of the $B$ meson. The flavor index contractions of the operators 
for the baryons are \cite{Arndt:2003vx}
\beq
(\bar{T} Y T)&=&\bar{T}^{ji} Y_i^\ell T_{\ell j},\nn \\
(\bar{S}^\mu Y S_\mu) &=& \bar{S}^{\mu, ji} Y_i^\ell S_{\mu,\ell j},\nn \\
(\bar{T} Y^\mu S_\mu) &=& \bar{T}^{ji} Y_i^{\mu,\ell} S_{\mu,\ell j}.
\eeq
The parameter $\Delta^{(B)}$ is the mass difference between the $S$ and $T$ fields with same light flavor indices,
\beq
\Delta^{(B)} = M_S-M_T,
\label{eq:m_shift_TS}
\eeq
which is of $O(\Lambda_{QCD})$ and does not vanish either in the chiral limit or in the heavy-quark limit.

In one-loop chiral calculations, 
the effects of SU$(6|3)$ flavor symmetry breaking appear through the mass differences of 
hadron fields: 
\beq
\mathcal{L}^{(\mathcal{\chi})}_\textrm{HH$\chi$PT}&=&
\tilde{\lambda}_1 \textrm{tr}_D [\bar{H}_i^{(\bar{b})} \calM^i_{\xi,j} H^{(\bar{b})j}]+
\lambda'_1 \textrm{tr}_D [\bar{H}_i^{(\bar{b})} H^{(\bar{b})i}] \textrm{str}(\calM_\xi)\nn \\
&&+\tilde{\lambda}_2 (\bar{S}^\mu \calM_\xi S_\mu)
+\lambda'_2 (\bar{S}^\mu S_\mu) \textrm{str}(\calM_\xi)
+\tilde{\lambda}_3 (\bar{T} \calM_\xi T)
+\lambda'_3 (\bar{T}T) \textrm{str} (\calM_\xi),
\eeq
where 
\beq
\calM_\xi = B_0 \left(\xi \calM^\dagger \xi + \xi^\dagger \calM \xi^\dagger \right).
\eeq
The terms proportional to $\lambda'$ cause a universal shift 
to the single-$b$ hadron masses, 
while the terms proportional to $\tilde{\lambda}$ attribute to the mass differences, 
\beq
\delta^{(M)}_{k,\ell}&=&M_{P_\ell}-M_{P_k}=M_{P_\ell^*}-M_{P_k^*}=2\tilde{\lambda}_1 B_0 (m_\ell-m_k),
\nn \\
\delta^{(B)}_{k,\ell}&=&M_{\calT_{\ell j}}-M_{\calT_{kj}}=M_{\calS_{\ell j}}-M_{\calS_{kj}}
=2\tilde{\lambda}_2 B_0 (m_\ell-m_k), 
\label{eq:m_shift_light}
\eeq
where $j$ represents the spectator quark and we assume $\tilde{\lambda}_2=\tilde{\lambda}_3$. 
We also consider the effects of heavy-quark spin symmetry breaking at $O(\Lambda_{QCD}/m_Q)$, 
\beq
\frac{\alpha}{m_b} \textrm{tr}_D \left(\bar{H}^{(\bar{b})}_i \sigma_{\mu\nu} H^{(\bar{b}),i} 
\sigma^{\mu\nu}\right),
\label{eq:lagrangian_Q}
\eeq
which vanishes in the heavy-quark limit. 
The $\alpha$ term provides the mass difference between the $P^*$ and $P$ mesons with same light flavor,
\beq
\Delta^{(M)}=M_{P^*_i}-M_{P_i}=-8\frac{\alpha}{m_b}.
\label{eq:m_shift_spin}
\eeq
In principle, there are also analogous heavy-quark spin symmetry breaking terms in the baryon sector,
resulting in mass differences between $B_{ij}$ and $B^{*\mu}_{ij}$ baryons in \Eq{S_field}. 
However, we neglect these mass differences 
which are numerically much smaller than $\Delta^{(B)}$ \cite{Aaltonen:2007ar}. 
If we redefine the field in which the heavy-light meson and $T$ baryon
containing $u$ or $d$ valence quarks are massless, 
the denominators of the propagators for other heavy-light mesons and single-$b$ baryons 
are shifted by the linear combinations of various mass differences in \Eq{m_shift_TS}, 
\Eq{m_shift_light}, and \Eq{m_shift_spin}. 

\section{$\Delta B=0$ Four-Quark Operators in Heavy Hadron Chiral Perturbation Theory}
\label{sec:four_quark_ops}

In the partially quenched theory, 
the four-quark operators in \Eq{four_quark_ops} transform as a left-flavor singlet, 
a $30$-dimensional representation, 
and a left-flavor octet, a $51$-dimensional representation, 
built from the tensor product of two fundamental representations of SU$(6|3)$. 
For lattice QCD, a more convenient form of the $\Delta B=0$ four-quark operators 
can be obtained by accounting for the graded light-flavor transformations\footnote{
In SU$(K+N|K)$ PQQCD with arbitrary $K$ (valence and ghost quarks) and 
$N$ (sea quarks), the singlet four-quark operators do not match those of unquenched QCD 
\cite{Golterman:2001qj}. 
When the number of valence quarks is equal to the number of sea quarks (i.e $K=N=3$ in our case); 
however, this mismatch disappears and the LO low-energy constants appearing in both theories 
are exactly the same. 
}:
\beq
\calO^{\lambda^{\textrm{PQ}}_k}=\textrm{str}\left(\lambda^{\textrm{PQ}}_k \calO\right)
+\delta^{\teye}\textrm{str}\left(\lambda^{\textrm{PQ}}_k\right) \textrm{str}\left(\calO\right),
\label{eq:FQ_Ops}
\eeq
with
\beq
\mathcal{O}^a_{~b}=\bar{b}\Gamma^{(1)}q_L^a \bar{q}_{L,b}\Gamma^{(2)} b,
\label{eq:FQ_ops}
\eeq
where
\beq
q^a_{L}=\frac{1-\gamma_5}{2}q^a,~\textrm{and}~~\bar{q}_{L,b}=\bar{q}_b\frac{1+\gamma_5}{2}.
\eeq
Here $\Gamma^{(1)}$ and $\Gamma^{(2)}$ are the appropriate spin and color matrices. 
Following the standard construction of the operators from heavy hadron fields, 
we can consider $\Gamma^{(1)}$ and $\Gamma^{(2)}$ as spurious fields. 
To make the operators invariant under SU$(2)$ heavy-quark spin 
and SU$(6|3)$ flavor transformations, the spurious fields must transform as
\beq
\Gamma^{(1)}\longrightarrow S_h\Gamma^{(1)} U_L^\dagger,~~~~~~~~~~\Gamma^{(2)}\longrightarrow U_L\Gamma^{(2)}S_h^\dagger.
\eeq
The symbol $\delta^{\teye}$ denotes the contribution of the  eye contractions, 
where the light quarks in the four-quark operators $\calO$ are contracted by themselves, 
while the symbol $\lambda^{\textrm{PQ}}$ denotes the partially quenched extension 
of the Gell-Mann matrix. 
It is convenient to classify the operators according to three types of quarks, 
(valence, sea, ghost), transformed under the chiral rotation of 
$\textrm{SU}(3)_{\textrm{val}}\bigotimes \textrm{SU}(3)_{\textrm{\tsea}}\bigotimes 
\textrm{SU}(3)_{\textrm{ghost}}$; 
the Gell-Mann matrices can be extended in a simple way, 
$\lambda^{\textrm{PQ}}_k=\lambda_k\bigotimes \textrm{diag} (1,\lambda^{\tsea},1)$, 
where $\lambda_0=\textrm{diag}(1,1,1)$ for the flavor singlet and $\lambda_3=\textrm{diag}(1,-1,0)$, 
$\lambda_8=\textrm{diag}(1,1,-2)$ for the flavor octet. 
Here, we have introduced the symbol $\lambda^{\tsea}$ to distinguish the disconnected (nonvalence) 
contributions from the connected (valence) ones, i.e. 
$\lambda^{\tsea}=1$ or $\lambda^{\tsea}=0$ depending whether the disconnected contributions are 
included or not. 

The LO operators in HH$\chi$PT relevant to the four-quark operators in \Eq{FQ_Ops} 
can be written as
\beq
\tilde{\calO}^{\lambda^{\textrm{PQ}}_k}=
\textrm{str} \left(\lambda_k^{\textrm{PQ}} \tilde{\calO}\right)
+\delta^{\teye}\textrm{str} \left(\lambda_k^{\textrm{PQ}}\right)  \textrm{str} 
\left(\tilde{\calO}^{(\textrm{bare})}\right), 
\label{eq:LO_four_quark_ops}
\eeq
where the operators with tildes $\tilde{\calO}$ for heavy-light mesons and single-$b$ baryons 
are defined in \Sec{four_quark_hmcpt} and \Sec{four_quark_hbcpt}, respectively.\footnote{
The operator $\tilde{O}^{(\textrm{bare})}$ is the tree-level operator of $\tilde{O}$, 
i.e. the four-quark operators are not dressed by Goldstone mesons. 
} 
At tree level in the chiral expansion, the first term corresponds to connected diagrams, while 
the second term corresponds to disconnected diagrams which 
are absent for the octet. 
Notice that, as will be seen in \Sec{matrix_elements}, the first term 
generates disconnected diagrams as well as connected ones at the next-to-leading-order (NLO) one-loop level. 

\subsection{$\Delta B=0$ four-quark operators for heavy-light $B$ mesons}
\label{sec:four_quark_hmcpt}

The most general bosonized form of the four-quark operators for heavy-light $B$ mesons in HH$\chi$PT is\footnote{
The field $H^{(b)}_i$ for mesons containing a $b$ quark and a light antiquark 
of flavor $i$ is obtained by applying the charge conjugation operation 
to the field $H^{(\bar{b})}_i$ in \eq{field_H}: 
$
H^{(b)}_i=\frac{1+\displaystyle{\not}v}{2}
\left(
P^{*(b)}_{i,\mu}\gamma^\mu-P^{(b)}_i\gamma_5
\right)$, 
which transforms under $S_h$ and $U$ as 
$
H^{(b)}_i \rightarrow S_h H^{(b)}_j U^{\dagger j}_{~~i}$.
Also, the conjugate field is defined as 
$\bar{H}^{(b)}_i=\gamma^0 H^{(b)\dagger}_i\gamma^0$. 
} 
\beq
\left(\tilde{\calO}^{\textrm{HM}}_i\right)^a_{~b}&=&\sum_x \left\lbrace \alpha^{(1)}_{i,x}
[(\xi \bar{H}^{(b)})^a\Gamma^{(1)}\Xi_x]
[\Gamma^{(2)}(H^{(b)}\xi^\dagger)_b\Xi'_x]
+\alpha^{(2)}_{i,x}
[(\xi \bar{H}^{(b)})^a\Gamma^{(1)}\Xi_x
\Gamma^{(2)}(H^{(b)}\xi^\dagger )_b\Xi'_x]\right. \nn \\
&&+\left. \alpha^{(3)}_{i,x}
[(\xi H^{(\bar{b})})^a\Gamma^{(1)}\Xi_x]
[\Gamma^{(2)}(\bar{H}^{(\bar{b})}\xi^\dagger)_b\Xi'_x]
+\alpha^{(4)}_{i,x}
[(\xi H^{(\bar{b})})^a\Gamma^{(1)}\Xi_x
\Gamma^{(2)}(\bar{H}^{(\bar{b})}\xi^\dagger )_b\Xi'_x]\right. \nn \\
&&+\left. \alpha^{(5)}_{i,x}
[(\xi \bar{H}^{(b)})^a(H^{(b)}\xi^\dagger)_b\Xi_x]
[\Gamma^{(2)}\Gamma^{(1)}\Xi'_x]
+\alpha^{(6)}_{i,x}
[(\xi \bar{H}^{(b)})^a(H^{(b)}\xi^\dagger)_b\Xi_x\Gamma^{(2)}\Gamma^{(1)}\Xi'_x]\right. \nn \\
&&+\left.\alpha^{(7)}_{i,x}
[(\xi H^{(\bar{b})})^a(\bar{H}^{(\bar{b})}\xi^\dagger)_b \Xi'_x]
[\Gamma^{(2)}\Gamma^{(1)}\Xi'_x]
+\alpha^{(8)}_{i,x}
[(\xi H^{(\bar{b})})^a(\bar{H}^{(\bar{b})}
\xi^\dagger)_b \Xi'_x\Gamma^{(2)}\Gamma^{(1)}\Xi'_x]\right\rbrace,\nn \\
\label{eq:op_hmchipt}
\eeq
where $i=1,2,3,4$, and $\Xi_x$ are $\Xi'_x$ are all possible pairs of Dirac structures.\footnote{
The Dirac structure $\Xi_x=1,~\gamma_\mu,~\sigma_{\mu\nu}$, 
and possible combinations with $\displaystyle{\not} v$ and $\gamma_5$.} 
The bracket notation $[~]$ represents the trace over the Dirac space. 
Any insertion of $\displaystyle{\not} v$ in the Dirac structures can be absorbed by the heavy meson fields, 
$\displaystyle{\not} v H(v)=H(v)$, while any insertion of $\gamma_5$ at most changes 
the sign of $\Gamma$. The HQET parity conservation 
and the contraction of Lorentz indices requires that $\Xi_x=\Xi'_x$. 
The single Dirac-trace terms can be rewritten by the double trace using the $4\times 4$ matrix identity
\beq
4[AB]=[A][B]+[\gamma_5 A][\gamma_5 B]+[A \gamma_\mu][\gamma^\mu B]
+[A\gamma_\mu \gamma_5][\gamma_5\gamma^\mu B]+\frac{1}{2}[A \sigma_{\mu\nu}][\sigma^{\mu\nu} ].
\eeq
Since $\Gamma^{(1)}$ and $\Gamma^{(2)}$ are left- and right-handed currents respectively, 
the last four terms in \Eq{op_hmchipt} vanish.
In summary, the operators in  $\tilde{\calO}^{\textrm{HM}}_{i}$ can be reduced to 
\beq
\left(\tilde{\calO}^{\textrm{HM}}_i\right)^a_{~b}&=&\sum_x \left\lbrace \bar{\alpha}_{i,x}
[(\xi \bar{H}^{(b)})^a\Gamma^{(1)}\Xi_x]
[\Gamma^{(2)}(H^{(b)}\xi^\dagger)_b\Xi_x]
\right.\nn \\
&&~~~~~~~~~~~~~~~~~~~~~~~~~~~~~~~~~~~~\left.
+\alpha_{i,x}
[(\xi H^{(\bar{b})})^a\Gamma^{(1)}\Xi_x]
[\Gamma^{(2)}(\bar{H}^{(\bar{b})}\xi^\dagger)_b\Xi_x]\right\rbrace, 
\eeq
where $\Xi_x \in \{1,\gamma_\nu,\sigma_{\nu\rho}\}$. 
After evaluating the trace over the Dirac space, we obtain 
\beq
\left(\tilde{\calO}^{\textrm{HM}}_i\right)^{a}_{~b}
&=&\bar{\beta}_{i,1}(\xi P^{(b)\dagger})^a(P^{(b)}\xi^\dagger)_b
+\bar{\beta}_{i,2}(\xi P^{*(b)\dagger}_{\mu})^a(P^{*(b),\mu}\xi^\dagger)_b\nn \\
&&~~~~~~~~~~~~~~~~~~~~~~~~+\beta_{i,1}(\xi P^{(\bar{b})})^a(P^{(\bar{b})\dagger}\xi^\dagger)_b
+\beta_{i,2}(\xi P^{*(\bar{b})}_{\mu})^a(P^{*(\bar{b})\dagger,\mu}\xi^\dagger))_b,
\label{eq:O_hmchipt}
\eeq
where $\beta$ and $\bar{\beta}$ are linear combinations of $\alpha_{i,x}$ and $\bar{\alpha}_{i,x}$, 
respectively. 
Explicitly we have, for $i=1,3$,
\beq
\beta_{i,1}(\bar{\beta}_{i,1})&=&4\alpha_{i,1}(\bar{\alpha}_{i,1})
+16\alpha_{i,\gamma_\nu}(\bar{\alpha}_{i,\gamma_\nu}), \nn \\
\beta_{i,2}(\bar{\beta}_{i,2})&=&4\alpha_{i,1}(\bar{\alpha}_{i,1}),
\label{eq:beta_13}
\eeq
and for $i=2,4$
\beq
\beta_{i,1}(\bar{\beta}_{i,1})&=&4\alpha_{i,1}(\bar{\alpha}_{i,1})
+4\alpha_{i,\gamma_\nu}(\bar{\alpha}_{i,\gamma_\nu}), \nn \\
\beta_{i,2}(\bar{\beta}_{i,2})&=&4\alpha_{i,\gamma_\nu}(\bar{\alpha}_{i,\gamma_\nu}).
\label{eq:beta_24}
\eeq
As shown above, the low-energy constants (LECs) for pseudoscalar and vector meson mixing processes 
are different in all cases, which is different from the results in Ref. \cite{Becirevic:2008us} 
where LECs for the cases of $i=2,4$ are the same. 
Since having a single LEC for the operators in an effective theory greatly 
simplifies chiral extrapolation of corresponding lattice data, 
it is important to confirm our result from another consideration. 
For this purpose, consider our four-fermion operators, $\mathcal{O}_{i,a}$, 
which appear in the HQET 
as 
\beq
\mathcal{O}^{\textrm{HQET}}_{i,a}=Q^\dagger \Gamma^{(1)} q_a \bar{q}_a \Gamma^{(2)} Q 
+ \tilde{Q} \Gamma^{(1)} q_a \bar{q}_a \Gamma^{(2)} \tilde{Q}^\dagger, 
\eeq
where $\Gamma$ are the appropriate Dirac and color structures, while 
$Q$ and $\tilde{Q}$ are the fields annihilating a heavy quark and heavy antiquark, respectively. 
Note that these fields do not create the corresponding antiparticles but their conjugate fields do.

We also consider the heavy-quark spin operator \cite{Savage:1989jx}  
\beq
S^3_Q=\epsilon^{ij3}[Q^\dagger \sigma_{ij} Q-\tilde{Q}\sigma_{ij}\tilde{Q}^\dagger],
\eeq
where $\sigma_{ij}=\frac{i}{2}[\gamma_i,\gamma_j]$ with $i,j=1,2,3$. 
Since the heavy-quark spin operator changes the spin of a heavy-light meson by one, 
the relation between LECs in the operator $\tilde{\calO}_{i}^{\textrm{HM}}$ 
can be seen from calculating the commutation (or anticommutation) of $S^3_Q$ 
and $\mathcal{O}_{i,0}^{\textrm{HQET}}$ acting on 
the pseudoscalar heavy-light meson state $|P\rangle$ containing a heavy-quark field $Q$. 
We find that
\beq
\lbrack S^3_Q, \mathcal{O}^{\textrm{HQET}}_{i,a}\rbrack |P\rangle \neq 0,
~~~\lbrace S^3_Q, \mathcal{O}^{\textrm{HQET}}_{i,a}\rbrace |P\rangle \neq 0.
\eeq
Similar results can be shown for the pseudoscalar heavy-light meson state $|\bar{P}\rangle$ 
containing a heavy antiquark field $\tilde{Q}$. 
This implies that the pseudoscalar and vector meson mixing processes via these operators are not simply related to each other,
and so they are accompanied by different LECs consistent with Eqs.~\ref{eq:O_hmchipt}-\ref{eq:beta_24}. 

\subsection{$\Delta B=0$ four-quark operators for single-$b$ baryons}
\label{sec:four_quark_hbcpt}

The most generalized form of four-quark operators for single-$b$ baryons in HH$\chi$PT is 
\beq
\left(\tilde{\calO}^{\textrm{HB}}_i\right)^a_{~b}&=&\sum_x \left\lbrace
\alpha'^{(1)}_{i,x} (\bar{S}_\mu \xi^{\dagger})^{aj} \Gamma^{(1)}\Xi^{\mu\nu}_x 
\Gamma^{(2)}(\xi S_\nu)_{bj}
+ \alpha'^{(2)}_{i,x} (\bar{S}_\mu \xi^{\dagger})^{aj} (\xi S_\nu)_{bj} 
\textrm{tr}_\textrm{D}[\Gamma^{(1)}\Xi^{\mu\nu}_x\Gamma^{(2)}]
\right. \nn \\
&&+\left. \alpha'^{(3)}_{i,x}(\bar{T}\xi^\dagger)^{aj}\Gamma^{(1)}\Xi_x \Gamma^{(2)}(\xi T)_{bj}
+ \alpha'^{(4)}_{i,x}(\bar{T}\xi^\dagger)^{aj}(\xi T)_{bj}
\textrm{tr}_\textrm{D}[\Gamma^{(1)}\Xi_x\Gamma^{(2)}]
\right\rbrace,
\label{eq:op_hbchipt}
\eeq
where $i=1,2,3,4$, and $\Xi$ and $\Xi^{\mu\nu}$ are all possible Dirac structures,
\beq
\Xi= 1,~\displaystyle{\not} v,~~~
\Xi^{\mu\nu}= g^{\mu\nu},~\displaystyle{\not} v g^{\mu\nu},~\sigma^{\mu\nu},~\displaystyle{\not} v \sigma^{\mu\nu},~\sigma^{\mu\nu} \displaystyle{\not} v, 
\eeq
and possible combinations with $\gamma^5$. Any insertion of $\gamma^5$ changes at most the overall sign. 
One may consider the terms $\bar{T}\Gamma^{(1)}\Xi^\mu_x \Gamma^{(2)}S_\mu$ and  
$\bar{T}S_\mu[\Gamma^{(1)}\Xi^\mu_x\Gamma^{(2)}]$ along with their conjugates which are 
invariant under SU$(2)$ heavy quark spin symmetry, 
where $\Xi^\mu=\gamma^\mu,~\displaystyle{\not} v \gamma^\mu,~\gamma^\mu \displaystyle{\not} v$. 
However, we exclude these terms because, along with $\lambda^{\textrm{PQ}}$, 
they violate the SU$(6|3)_{\textrm{L+R}}$ flavor symmetry. 
By evaluating the Dirac matrices we obtain 
\beq
\left(\tilde{\calO}^{\textrm{HB}}_{i}\right)^a_{~b}=\left\lbrace
\beta'_{i,1} (\bar{B} \xi^\dagger)^{aj}(\xi B)_{bj}
+\beta'_{i,2} (\bar{B}^*_\mu \xi^\dagger)^{aj}(\xi B^{* \mu})_{bj}
+\beta'_{i,3} (\bar{T} \xi^\dagger)^{aj}(\xi T)_{bj}\right\rbrace,
\label{eq:O_hhchipt}
\eeq
where $\beta'$'s are linear combinations of $\alpha'$'s. 
The explicit values of $\beta'$'s are, for $i=1,3$, 
\beq
\beta'_{i,1}=4 \alpha'^{(1)}_{i,\displaystyle{\not}v g^{\mu\nu}}
-8i(\alpha'^{(1)}_{i,\displaystyle{\not}v \sigma^{\mu\nu}}
+\alpha'^{(1)}_{i,\sigma^{\mu\nu} \displaystyle{\not} v}),~
\beta'_{i,2}=-\frac{1}{2}\beta'_{i,1}-2 \alpha'^{(1)}_{i,\displaystyle{\not}v g^{\mu\nu}},~
\beta'_{i,3}=-4 \alpha'^{(3)}_{i,\displaystyle{\not}v},
\eeq
and for $i=2,4$, 
\beq
\beta'_{i,1}=-2 \alpha'^{(1)}_{i,\displaystyle{\not}v g^{\mu\nu}}
-4i(\alpha'^{(1)}_{i,\displaystyle{\not}v \sigma^{\mu\nu}}
+\alpha'^{(1)}_{i,\sigma^{\mu\nu} \displaystyle{\not}v}),~
\beta'_{i,2}=\frac{1}{2}\beta'_{i,1}+3 \alpha'^{(1)}_{i,\displaystyle{\not}v g^{\mu\nu}},~
\beta'_{i,3}=2 \alpha'^{(3)}_{i,\displaystyle{\not}v},
\eeq
Unfortunately, this result implies that none of the low-energy constants 
are related to each other, and thus all of them must be considered as independent parameters 
in fitting lattice QCD data.  

\section{Matrix Elements of $\Delta B=0$ Operators}
\label{sec:matrix_elements}

The generic forms of the matrix elements of $\Delta B=0$ four-quark operators 
to the NLO in SU$(6|3)$ partially quenched HH$\chi$PT 
can be written as
\beq
\langle B|\calO^\tlam|B\rangle
=(1+3\lambda^{\tsea}\delta^{\teye}\delta_{k0})
\left[C_M^k \beta_1(1+\calW_{B})+\beta_{2} \calQ^k_{B}\right]
+\beta_{1} \calT^k_{B}
+\textrm{analytical terms},
\label{eq:mat_elements_B_meson}
\eeq
for $B$ mesons\footnote{
More precisely, $|B\rangle$ denotes the state annihilating a $B$ meson 
containing an anti-$b$ quark and a light quark, 
where we use the standard notation, e.g. $B_u=B^+=\bar{b}u$. 
It is straightforward to extend our calculations to the $\Delta B=0$ 
matrix elements involving mesons containing a $b$ quark and a light antiquark. 
} and 
\beq
\langle T|\calO^\tlam|T\rangle
&=&(1+3\lambda^{\tsea}\delta^{\teye}\delta_{k0})
\left[C_T^k \beta'_{3} (1+\calW_{T}) 
+(-\beta'_{1}+2\beta'_{2}) \calQ^k_{T}\right]
+\beta'_{3}\calT^k_{T}
\nn \\
&&+\textrm{analytical terms},
\nn \\
\langle S|\calO^\tlam|S\rangle
&=& (1+3\lambda^{\tsea}\delta^{\teye}\delta_{k0})
\left[C_S^k\beta'_{1} (1+\calW_{S})
+(-2 \beta'_{1}+\beta'_{2}) \calQ^k_{S_S}
+\beta'_{3} \calQ^k_{S_T}\right]
+\beta'_{1}\calT^k_{S}
\nn \\
&&+\textrm{analytical terms},
\label{eq:mat_elements_B_baryon}
\eeq
for $T$ and $S$ baryons. 
The equality symbol means the matching between PQQCD and the chiral effective theory. 
The coefficients $C_M^0=C_T^0=C_S^0=1$ 
for all external single-$b$ hadrons, while those with $k=3, 8$ are given in \Tab{tlam_meson} 
for $B$ mesons and in \Tab{tlam_baryon} for $T$ and $S$ baryons. 

\begin{table}
\caption{%
\label{tab:tlam_meson}
Coefficients $C_M^k$ in \Eq{mat_elements_B_meson}.
}
\begin{center}
\begin{tabular}{c|ccc}
&\multicolumn{3}{c}{$C^{k}_M$}\\
\cline{2-4}
&~~~$B_u$~~~&~~~$B_d$~~~&~~~$B_s$~~~\\
\hline
~~~$k=3$~~~ & $1$ & $-1$ & $0$ \\
$k=8$ & $1$  & $1$ & $-2$ \\
\hline\hline
\end{tabular}
\end{center}
\end{table}

\begin{table}
\caption{%
\label{tab:tlam_baryon}
Coefficients $C_T^k$ and $C_S^k$ in \Eq{mat_elements_B_baryon}.
}
\begin{center}
\begin{tabular}{c|ccc|cccccc}
&\multicolumn{3}{c|}{$C^{k}_T$}&\multicolumn{6}{c}{$C^{k}_S$}\\
\cline{2-10}
&~~~$\Lambda_b$~~~&~~~$\Xi^{+\frac{1}{2}}_b$~~~&~~~$\Xi^{-\frac{1}{2}}_b$~~~
&~~~$\Sigma_b^0$~~~&~~~$\Sigma_b^+$~~~&~~~$\Sigma_b^-$~~~&~~~$\Xi'^{+\frac{1}{2}}_b$~~~
&~~~$\Xi'^{-\frac{1}{2}}_b$~~~&~~~$\Omega_b$~~~\\
\hline
~~~$k=3$~~~ & 0 & $\frac{1}{2}$ & $-\frac{1}{2}$ 
& 0 & $1$ & $-1$ & $\frac{1}{2}$ & $-\frac{1}{2}$ & 0\\
$k=8$ & $1$  & $-\frac{1}{2}$ & $-\frac{1}{2}$ 
& $1$ & $1$ & $1$ 
& $-\frac{1}{2}$ & $-\frac{1}{2}$ & $-2$\\
\hline\hline
\end{tabular}
\end{center}
\end{table}

\begin{figure}
\centerline{
\xy
\xyimport(1,1){
\includegraphics[width=0.3\textwidth]{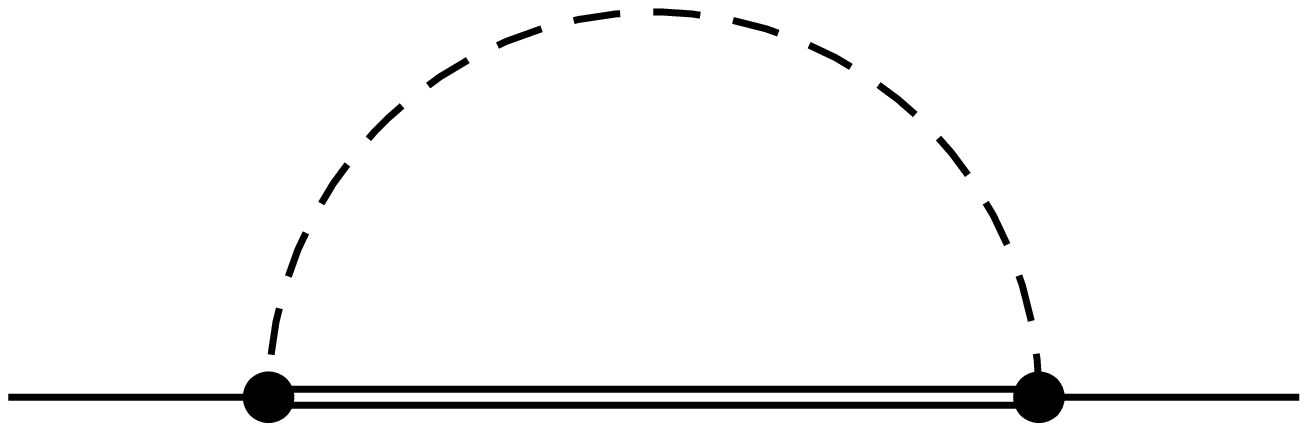}
\includegraphics[width=0.3\textwidth]{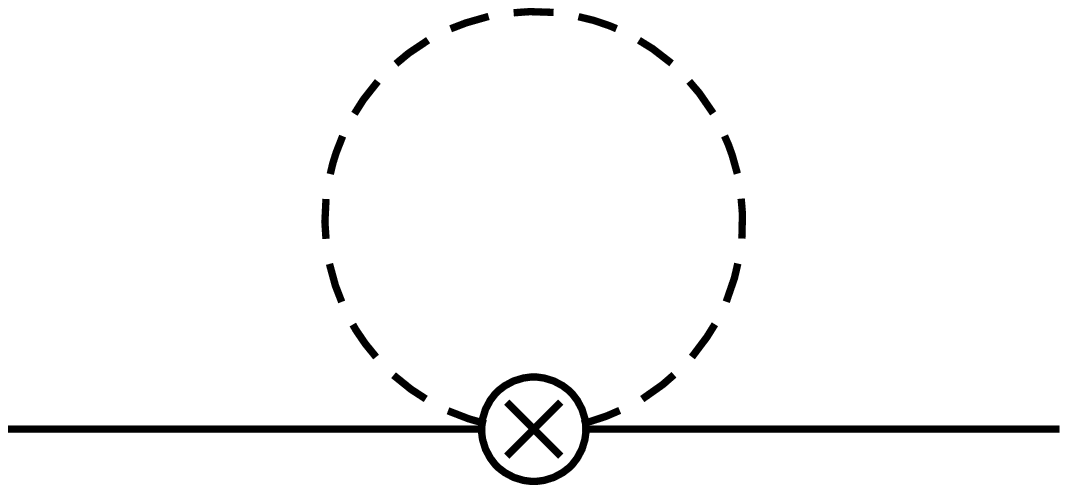}
\includegraphics[width=0.3\textwidth]{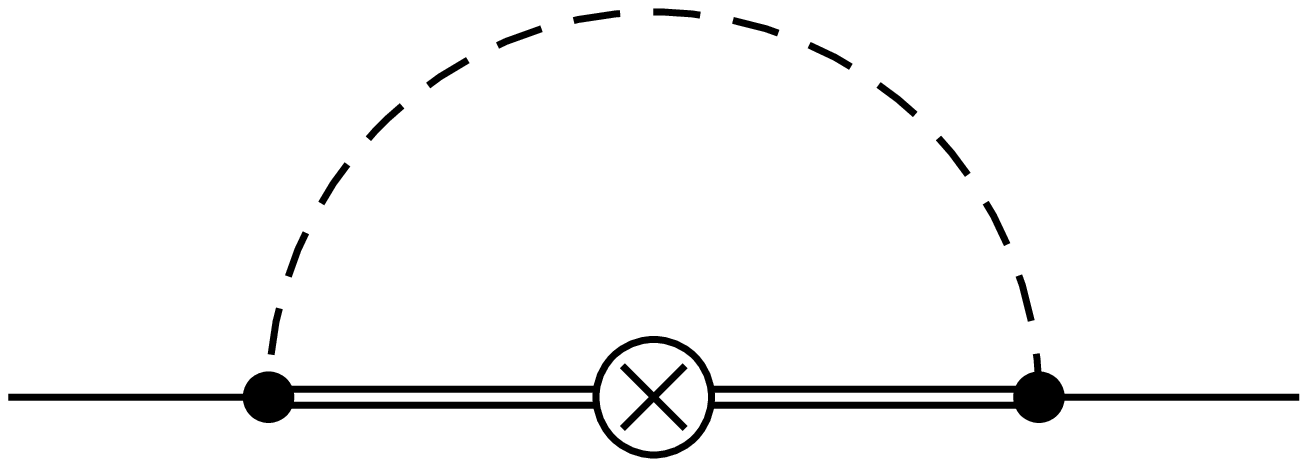}
}
\endxy
}
\caption{
One-loop diagrams contributing to the matrix elements of $\Delta B=0$ four-quark operators. 
The single and double solid lines correspond to external and internal heavy hadrons, 
such as $B$, $B^*$ mesons or $T$, $S$ baryons, respectively. 
The dashed lines are the Goldstone meson propagators including possible hairpin interactions. 
The crossed circles denote the LO four-quark operators in HH$\chi$PT in \Eq{LO_four_quark_ops}, 
while the filled circles denote 
the axial couplings in the chiral Lagrangian in \Eq{lagrangian_hhcpt}. 
Diagrams from left to right are the wavefunction renormalization, tadpole- and sunset-type operator 
renormalizations, respectively. 
}
\label{fig:one_loop}%
\end{figure}

There are three types of nonanalytic one-loop contributions, 
the wavefunction renormalization resulting from the self-energy diagrams $(\calW)$, 
tadpole $(\calT)$, and sunset $(\calQ)$ diagrams, as depicted in \Fig{one_loop}. 
We perform these calculations both at infinite volume and on a torus of length $L$ 
in each of three spatial directions (the temporal extent is assumed to be infinite), 
where the results are presented in \Sec{one_loop_result_B_meson} and \Sec{one_loop_result_B_baryon} below. 
Because the $\Delta B=0$ four-quark operators in HH$\chi$PT have the same forms 
for any values of $i$ as in 
\Eq{O_hmchipt} and \Eq{O_hhchipt}, the one-loop chiral corrections are independent on the structure 
of the $\Delta B=0$ operators in \Eq{four_quark_ops} and thus we omit the subscript $i$ throughout 
this section for convenience. 

The analytic terms include the NLO contributions of chiral symmetry breaking effects $\sim m_\phi^2$ 
and heavy quark symmetry breaking effects $\sim\Lambda_{QCD}/m_b$. 
In the chiral expansion, in particular, the former contributions play a role of 
counterterms which are necessary to renormalize the one-loop contributions. 
There are five corresponding NLO operators, str$\left(\lambda^{\textrm{PQ}}_k \calM_\xi \tilde{\calO}\right)$, 
str$\left(\lambda^{\textrm{PQ}}_k \tilde{\calO}\right)$str$\left(\calM_\xi\right)$, 
str$\left(\lambda^{\textrm{PQ}}_k \calM_\xi\right)$str$\left(\tilde{\calO}\right)$, 
str$\left(\lambda^{\textrm{PQ}}_k\right)$str$\left(\tilde{O}\calM_\xi\right)$, 
and $\left(\lambda^{\textrm{PQ}}_k\right)$str$\left(\tilde{O}\right)$str$\left(\calM_\xi\right)$, 
accompanying with unknown low-energy constants. 
Notice that in the isospin and QCD limit the last three operators 
for $\lambda^{\textrm{PQ}}_3$ vanish, 
while those for $\lambda^{\textrm{PQ}}_8$ are proportional to $m_K^2-m_\pi^2$ originate 
from the flavor SU$(3)$ symmetry breaking.

\subsection{One-loop contributions for $B$ Meson}
\label{sec:one_loop_result_B_meson}

First, we study the one-loop contributions for $B$ mesons in \Eq{mat_elements_B_meson}. 
To take account of the flavor SU$(3)$ breaking effects from 
both the Goldstone masses and the heavy-meson spectrum, 
we investigate the structure of the one-loop diagrams by 
analyzing the quark-flavor flow picture \cite{Sharpe:1992ft}. 
To do this, we define the rules as below. 
\begin{itemize}
\item Each flavor flow line in a one-loop diagram has a direction and a flavor index: 
the flow along the direction means a quark with that flavor, while the flow against the direction 
means its antiquark. We drop the flavor index of a $b$ quark because of 
its irrelevance to our discussion. 
\item The ``tilded" coefficients are for the hairpin 
contributions from the flavor-neutral mesons. 
\item The ``bar" notation in the coefficients 
denotes the flavor-disconnected diagrams appearing in the chiral expansion, 
which are distinguished from the eye contraction in \Eq{LO_four_quark_ops}. 
\end{itemize}

\begin{figure}
\centerline{
\begin{tabular}{cc}
\subfloat[]{
\xy
\xyimport(1100,788){
\includegraphics[width=0.45\textwidth]{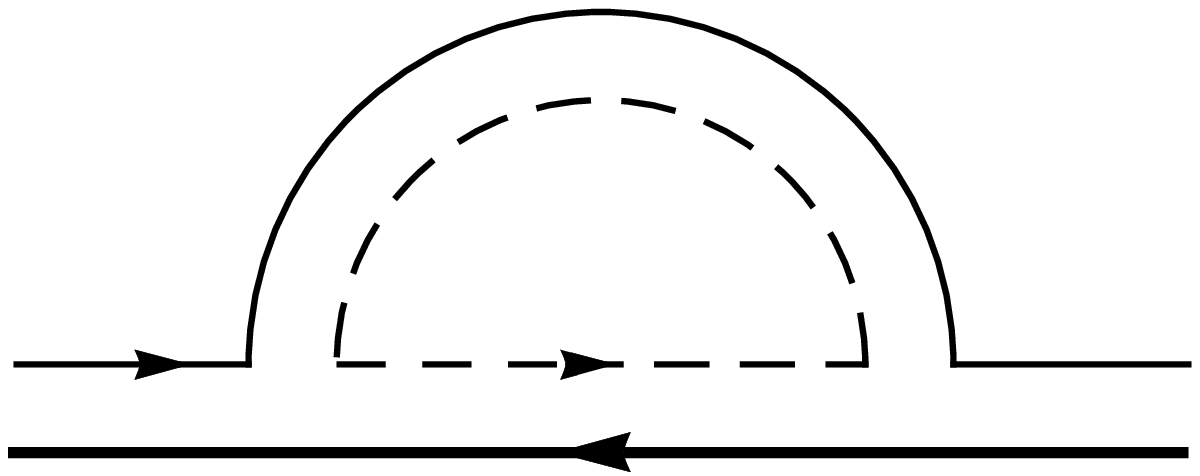}
\label{fig:selfenergy_B_meson}
}
(100,200)*{a};
(1000,200)*{a};
\endxy
}
\hspace{.1in}
&
\hspace{.1in}
\subfloat[]{
\xy
\xyimport(1100,788){
\label{fig:selfenergy_hairpin_B_meson}
\includegraphics[width=0.45\textwidth]{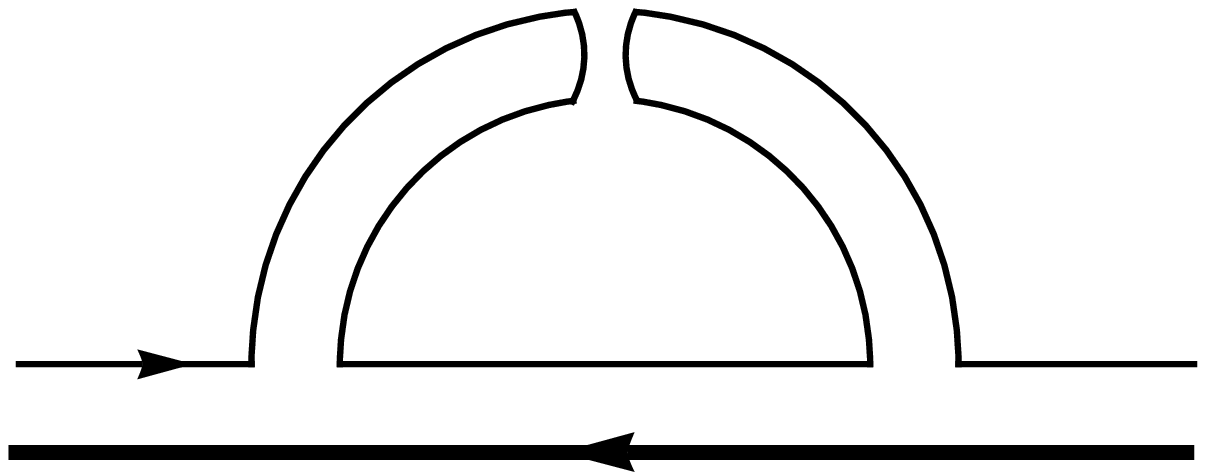}
}
(100,200)*{a};
(1000,200)*{a};
\endxy
}
\end{tabular}
}
\caption{%
\label{fig:self_energy_meson}
Self-energy diagrams contributing to the matrix elements containing $B$ mesons. 
The solid thick and thin lines denote the anti-$b$ quark and the valence light quark, respectively, 
while the dashed thin line denotes the sea light quark. 
Diagram (b) is the hairpin structure. 
}
\end{figure}

Following the above rules and the quark flow picture in \Fig{self_energy_meson}, 
the wavefunction renormalization contributions can be written as \cite{Detmold:2006gh}
\beq
\calW_{B_a}&=&\frac{ig_1^2}{f^2}\left[6\calH(M_{a,j},\Delta^{(M)}+\delta^{(M)}_{a,j})
+3\calH(M_{a,r},\Delta^{(M)}+\delta^{(M)}_{a,r})-\tilde{\calH}(M_{a,a},\Delta^{(M)})
\right],
\eeq
where $a$ runs over the valence light-quark flavors. 
The nonanalytic functions $\calH$ and $\tilde{\calH}$ 
are defined in the \Appendix{integrals_and_sums}, while 
the mass parameters $M_{a,b}$, $\Delta^{(M)}$ and $\delta^{(M)}_{a,b}$ are defined in \Sec{pqhhcpt}.

\begin{figure}
\centerline{
\begin{tabular}{cc}
\subfloat[]{
\label{fig:sunset_B_meson}
\xy
\xyimport(1100,788){
\includegraphics[width=0.45\textwidth]{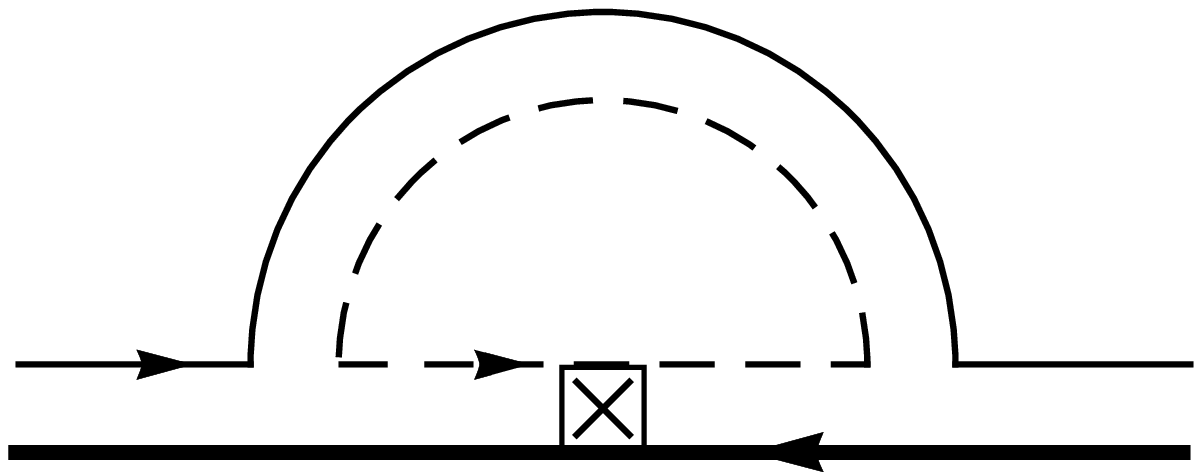}
}
(100,200)*{a};
(1000,200)*{a};
(460,300)*{a'};
\endxy
}
\hspace{.1in}
&
\hspace{.1in}
\subfloat[]{
\label{fig:sunset_hairpin_B_meson}
\xy
\xyimport(1100,788){
\includegraphics[width=0.45\textwidth]{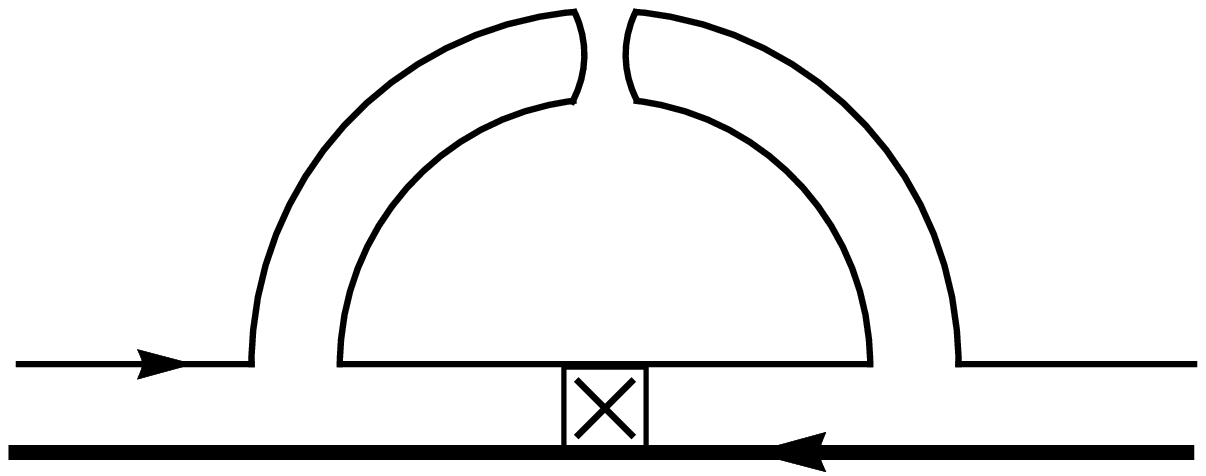}
}
(100,200)*{a};
(1000,200)*{a};
\endxy
}
\end{tabular}
}
\caption{%
\label{fig:sunset_meson}
Sunset diagrams contributing to the matrix elements containing $B$ mesons. 
The solid thick and thin lines denote the anti-$b$ quark and the valence light quark, respectively. 
The crossed square is the operator $\tilde{\calO}^\tlam$. 
Diagram (b) is the hairpin structure. 
}
\end{figure}

The loop contributions for tadpole integrals are completely determined 
by the structure of the operators $\tilde{\calO}^{\textrm{HM}}$ in \Eq{O_hmchipt}, 
where the flavor SU$(3)$ breaking effects arise solely from 
the Goldstone masses. As a result, we obtain
\beq
\calT^k_{B}&=&\frac{i}{f^2}\sum_\phi (x_{\phi_{ab}}^{B,k}
+\lambda^{\tsea}\bar{x}_{\phi_{ab}}^{B,k}) \calI(M_{a,b}),
\label{eq:tadpole_meson}
\eeq
where the summation runs over all possible Goldstone mesons 
in SU$(6|3)$ PQ$\chi$PT. 
The nonanalytic function $I$ is defined in \Appendix{integrals_and_sums} 
and the coefficients $x$ and $\bar{x}$ are given in \Tab{tadpole_meson} in \Appendix{coefficients_B_meson}. 
Notice that all hairpin diagrams 
cancel out and do not contribute to any one-loop tadpole corrections.

The quark flow picture for sunset diagrams is presented in \Fig{sunset_meson}, 
where the crossed square represents the operator $\tilde{\calO}^\tlam$ 
including both the eye and noneye contraction in \Eq{LO_four_quark_ops}.
The hairpin diagrams for the noneye contraction give rise to the 
connected contribution, while the others give rise to the disconnected contribution. 
The contributions from the sunset diagrams are then summarized as 
\beq
\calQ^k_{B}&=&\frac{ig_1^2}{f^2}
\left[\sum_\phi \lambda^{\tsea}\bar{y}_{\phi_{ab}}^{B,k} \calH(M_{a,b},\Delta^{(M)}+\delta^{(M)}_{a,b})
-\sum_{\phi\phi'}\tilde{y}_{\phi_{aa}\phi'_{bb}}^{B,k}
\tilde{\calH}(M_{a,b},\Delta^{(M)})\right],
\label{eq:sunset_meson}
\eeq
where the second term sums over all pairs of flavor-neutral states in the valence-valence sector, 
i.e. $\phi\phi'$ runs over $\eta_u\eta_u$, $\eta_u\eta_s$, and $\eta_s\eta_s$. 
The coefficients are given in \Tab{sunset_meson} in \Appendix{coefficients_B_meson}.

\subsection{One-loop contributions for single-$b$ Baryon}
\label{sec:one_loop_result_B_baryon}

\begin{figure}
\centerline{
\begin{tabular}{cc}
\subfloat[]{
\xy
\xyimport(1100,788){
\includegraphics[width=0.45\textwidth]{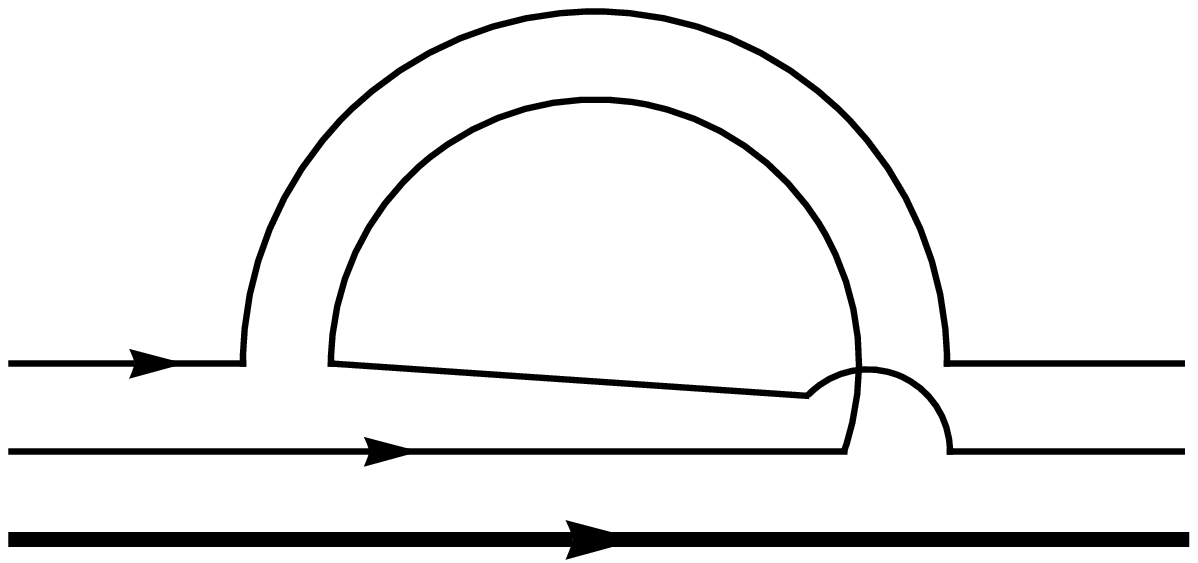}
}
(100,280)*{a};
(1000,280)*{a};
(100,150)*{b};
(1000,150)*{b};
\endxy
}
\hspace{.1in}
&
\hspace{.1in}
\subfloat[]{
\xy
\xyimport(1100,788){
\includegraphics[width=0.45\textwidth]{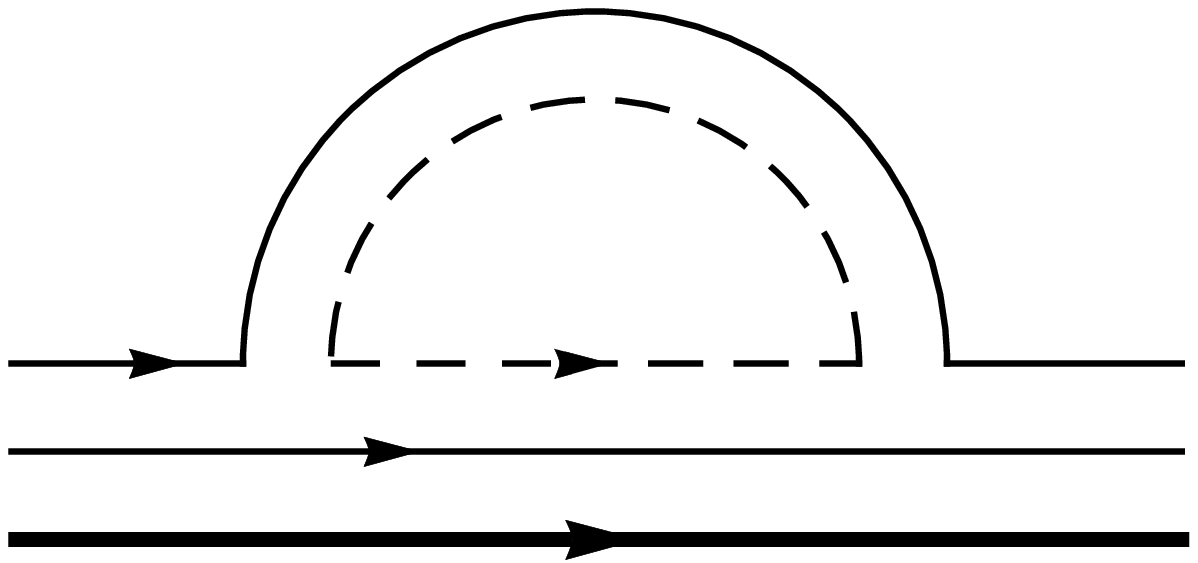}
}
(100,280)*{a};
(1000,280)*{a};
(100,150)*{b};
(1000,150)*{b};
(460,350)*{a'};
\endxy
}
\\
\subfloat[]{
\xy
\xyimport(1100,788){
\includegraphics[width=0.45\textwidth]{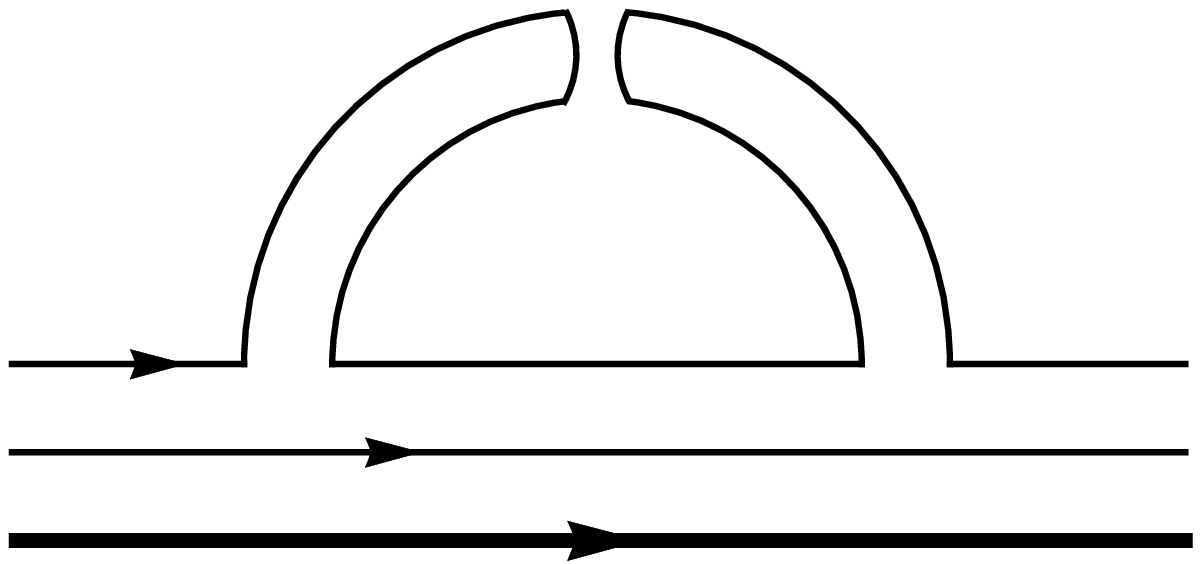}
}
(100,280)*{a};
(1000,280)*{a};
(100,150)*{b};
(1000,150)*{b};
\endxy
}
\hspace{.1in}
&
\hspace{.1in}
\subfloat[]{
\xy
\xyimport(1100,788){
\includegraphics[width=0.45\textwidth]{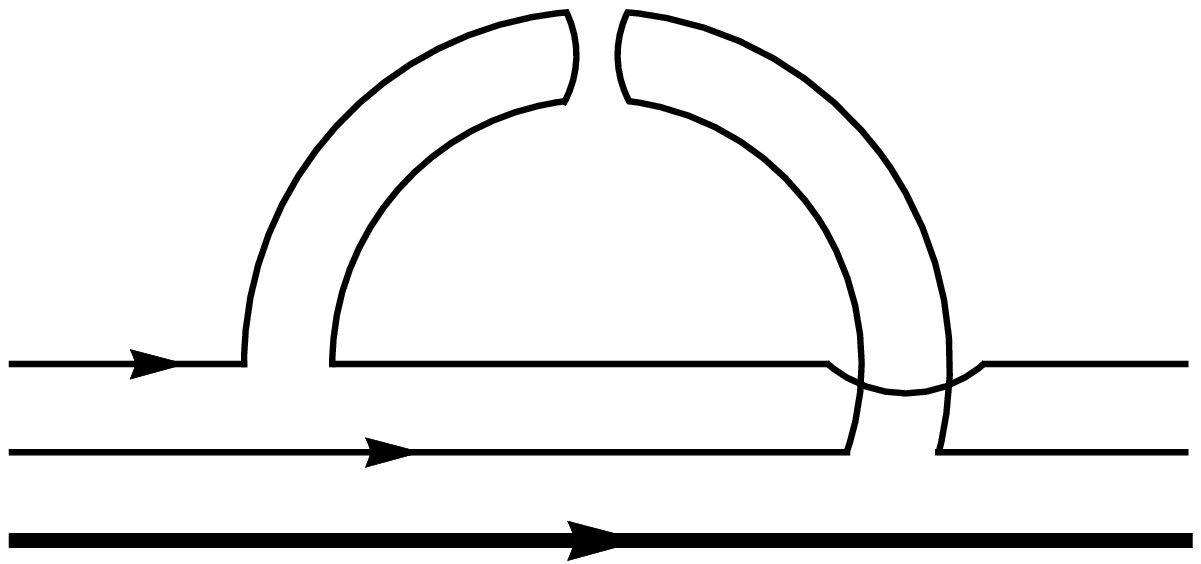}
}
(100,280)*{a};
(1000,280)*{a};
(100,150)*{b};
(1000,150)*{b};
\endxy
}
\end{tabular}
}
\caption{%
\label{fig:self_energy_baryon}
Self-energy diagrams contributing to the matrix elements containing single-$b$ baryons. 
The solid thick and thin lines denote the anti-$b$ quark and the valence light quark, respectively, 
while the dashed thin line denotes the sea light quark. 
Diagram (c) and (b) are the hairpin structures. 
The diagrams including the internal $S$ baryons lead to terms 
multiplied by $w$ and $\tilde{w}$, 
while those including the internal $T$ baryons lead to terms 
multiplied by $w'$ and $\tilde{w}'$ in \Eq{wavefunction_baryon}, 
where the tilded coefficients denote the hairpin structure. 
The diagrams obtained by interchanging the flavor indices $a$ and $b$ 
, not shown in this figure, 
have also been taken into account in our results in \Tab{wavefunction_baryon}. 
}
\end{figure}

To discuss the structure of the single-$b$ baryon one-loop diagrams within 
the quark-flavor flow picture, 
we introduce one more rule:
\begin{itemize}
\item The ``primed" coefficients are for the one-loop diagrams involving 
the internal $T$ baryon, while the unprimed coefficients 
for the diagrams involving the internal $S$ baryon. 
\end{itemize}
In \Fig{self_energy_baryon}, we show the quark-flavor flow picture for 
the baryon self-energy diagrams. 
Following the above rule and the rules defined in \Sec{one_loop_result_B_meson}, 
we obtain the contributions from the wavefunction renormalization 
for $T$ and $S$ baryons \cite{Tiburzi:2004kd, Detmold:2011rb},
\beq
\calW_T&=&\frac{ig_3^2}{f^2}\left[
\sum_\phi w^T_{\phi_{ab}}\calH(M_{a,b},\Delta^{(B)}+\delta^{(B)}_{a,b})-
\sum_{\phi\phi'}\tilde{w}^T_{\phi_{aa}\phi'_{bb}}\tilde{\calH}(M_{a,b},\Delta^{(B)})\right],
\nn \\
\calW_S&=&
\frac{ig_2^2}{f^2}\left[
\sum_\phi w^S_{\phi_{ab}}\calH(M_{a,b},\delta^{(B)}_{a,b})-
\sum_{\phi\phi'}\tilde{w}^S_{\phi_{aa}\phi'_{bb}}\tilde{\calH}(M_{a,b},0)\right]
\nn \\
&&
+\frac{ig_3^2}{f^2}\left[
\sum_\phi w'^S_{\phi_{ab}}\calH(M_{a,b},-\Delta^{(B)}+\delta^{(B)}_{a,b})-
\sum_{\phi\phi'}\tilde{w'}^S_{\phi_{aa}\phi'_{bb}}\tilde{\calH}(M_{a,b},-\Delta^{(B)})\right], 
\label{eq:wavefunction_baryon}
\eeq
where the summations are over the Goldstone mesons in SU$(6|3)$ PQ$\chi$PT.  
The coefficients $w, w', \tilde{w}$, and $\tilde{w}'$ are presented in 
\Tab{wavefunction_baryon} in \Appendix{coefficients_B_baryon}, while 
the mass parameters $\Delta^{(B)}$ and $\delta^{(B)}_{a,b}$ are defined above in \Sec{pqhhcpt}.

A single-$b$ baryon carries two light-flavor quarks, 
where one of them is contracted with the four-quark operators, 
while the other is the spectator quark. Therefore, the situation 
for the baryon tadpole and sunset diagrams is more involved compared to those 
for the meson diagrams. 
We first obtain the contributions from tadpole diagrams,
\beq
\calT^k_{T (S)}=\frac{i}{f^2}\sum_\phi (x^{T (S),k}_{\phi_{ab}}
+\lambda^{\tsea}\bar{x}^{T (S),k}_{\phi_{ab}}) 
\calI(M_{a,b}),
\label{eq:tadpole_baryon}
\eeq
where the coefficients $x^{T(S)}$ and $\bar{x}^{T(S)}$ are given in \Tab{tadpole_baryon} 
in \Appendix{coefficients_B_baryon}. 
Like the case of $B$ mesons, all hairpin diagrams 
cancel out and do not contribute to any one-loop tadpole corrections.

\begin{figure}
\centerline{
\begin{tabular}{cc}
\subfloat[]{
\xy
\xyimport(1100,788){
\includegraphics[width=0.45\textwidth]{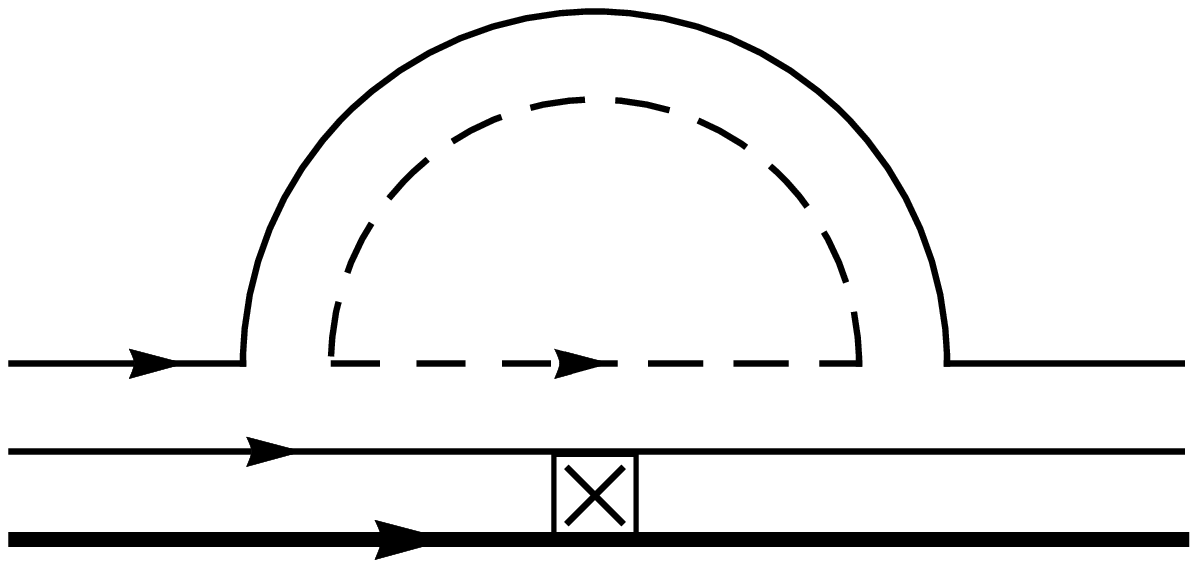}
}
(100,280)*{a};
(1000,280)*{a};
(460,350)*{a'};
(100,150)*{b};
(1000,150)*{b};
\endxy
}
&
\subfloat[]{
\xy
\xyimport(1100,788){
\includegraphics[width=0.45\textwidth]{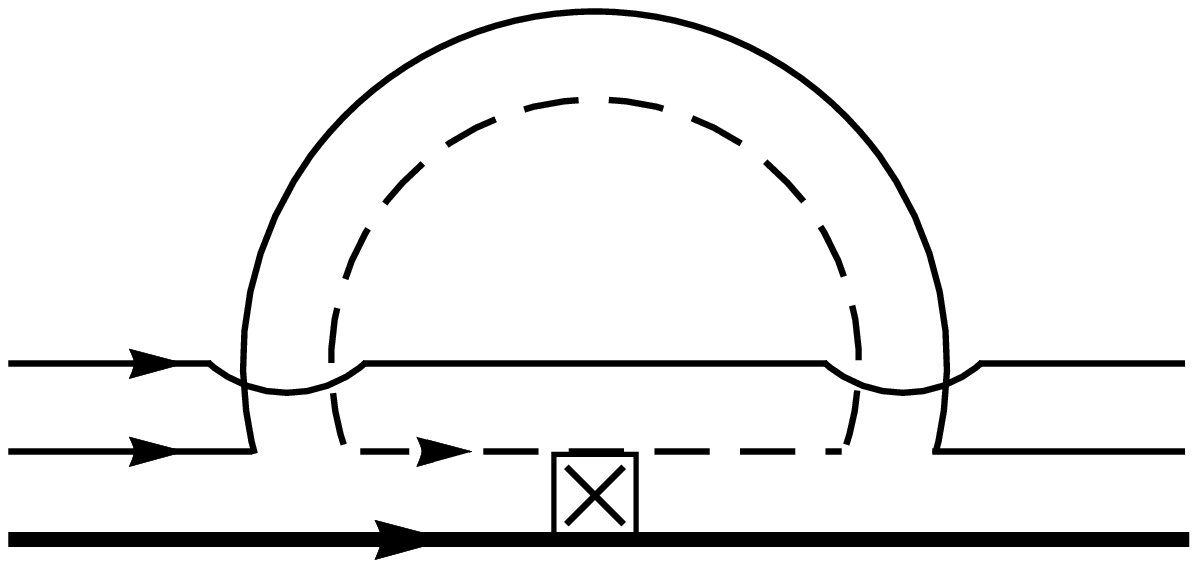}
}
(100,280)*{b};
(1000,280)*{b};
(480,220)*{a'};
(100,150)*{a};
(1000,150)*{a};
\endxy
\label{fig:sunset_baryon_2}
}
\\
\subfloat[]{
\xy
\xyimport(1100,788){
\includegraphics[width=0.45\textwidth]{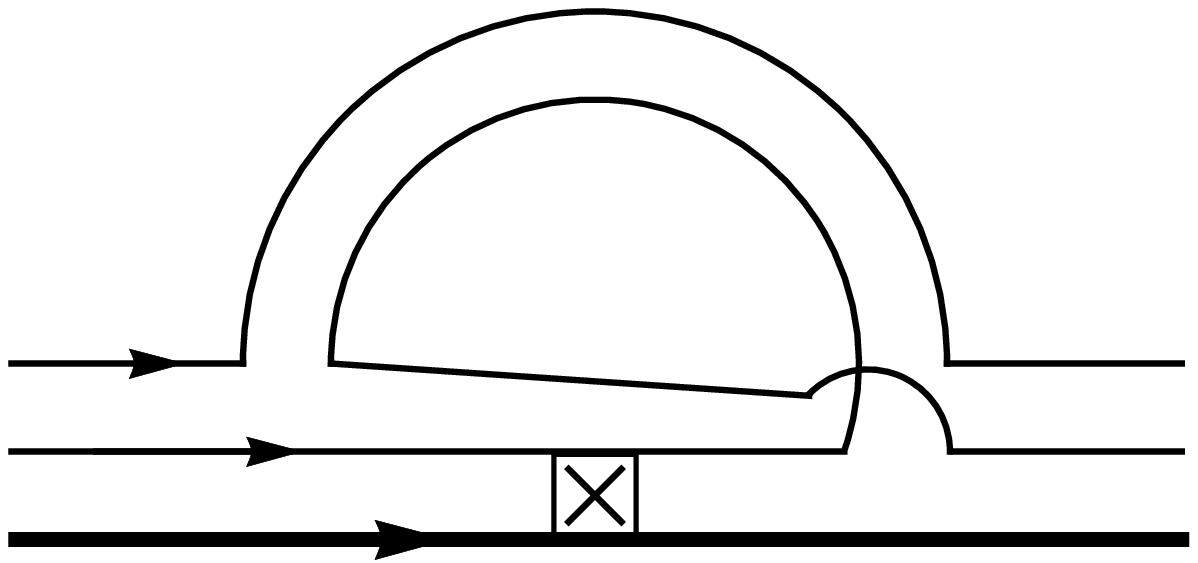}
}
(100,280)*{a};
(1000,280)*{a};
(100,150)*{b};
(1000,150)*{b};
\endxy
}
&
\subfloat[]{
\xy
\xyimport(1100,788){
\includegraphics[width=0.45\textwidth]{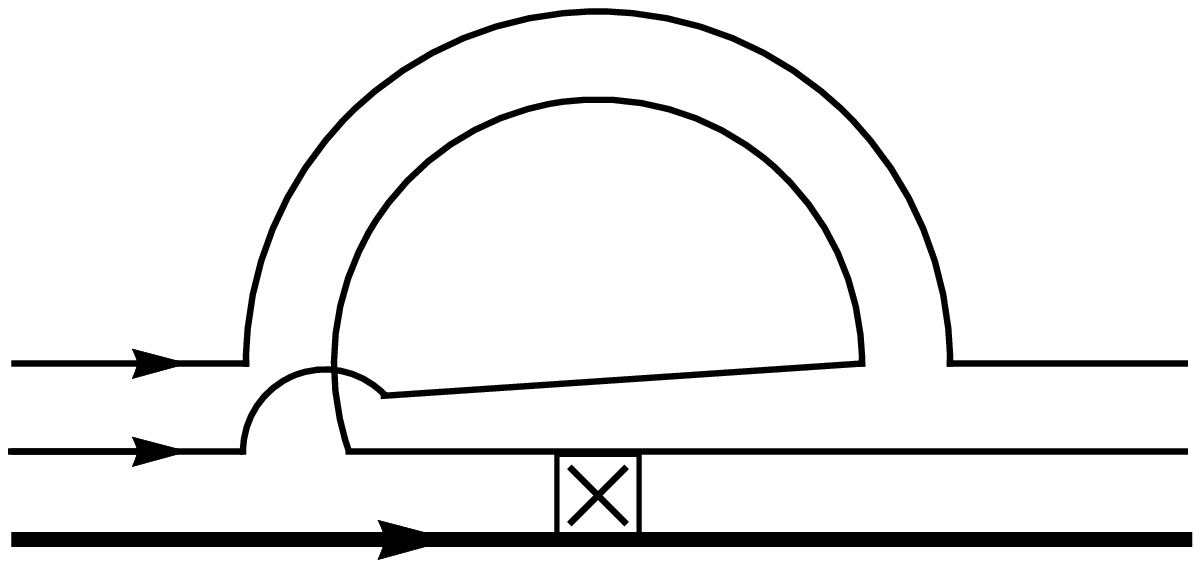}
}
(100,280)*{a};
(1000,280)*{a};
(100,150)*{b};
(1000,150)*{b};
\endxy
}
\end{tabular}
}
\caption{%
\label{fig:sunset_baryon}
Sunset diagrams contributing to the $\Delta B=0$ matrix element of single-$b$ baryons without 
the hairpin structure. 
The solid thick and thin lines denote the anti-$b$ quark and the valence light quark, respectively, 
while the dashed thin line denotes the sea light quark. 
The crossed square is the operator $\tilde{\calO}^\tlam$. 
Diagram (b) contributes to the $\bar{y}$ terms, 
while the diagrams including the internal $S$ and $T$ baryons lead to terms 
multiplied by $y$ and $y'$ in \Eq{sunset_baryon}, 
respectively. 
The diagrams obtained by interchanging the flavor indices $a$ and $b$ 
, not shown in this figure, 
have also been taken into account in our results in \Tab{sunset_T_baryon}, \Tab{sunset_S_baryon_1} 
and \Tab{sunset_S_baryon_2}. 
}
\end{figure}

\begin{figure}
\centerline{
\begin{tabular}{cc}
\subfloat[]{
\xy
\xyimport(1100,788){
\includegraphics[width=0.45\textwidth]{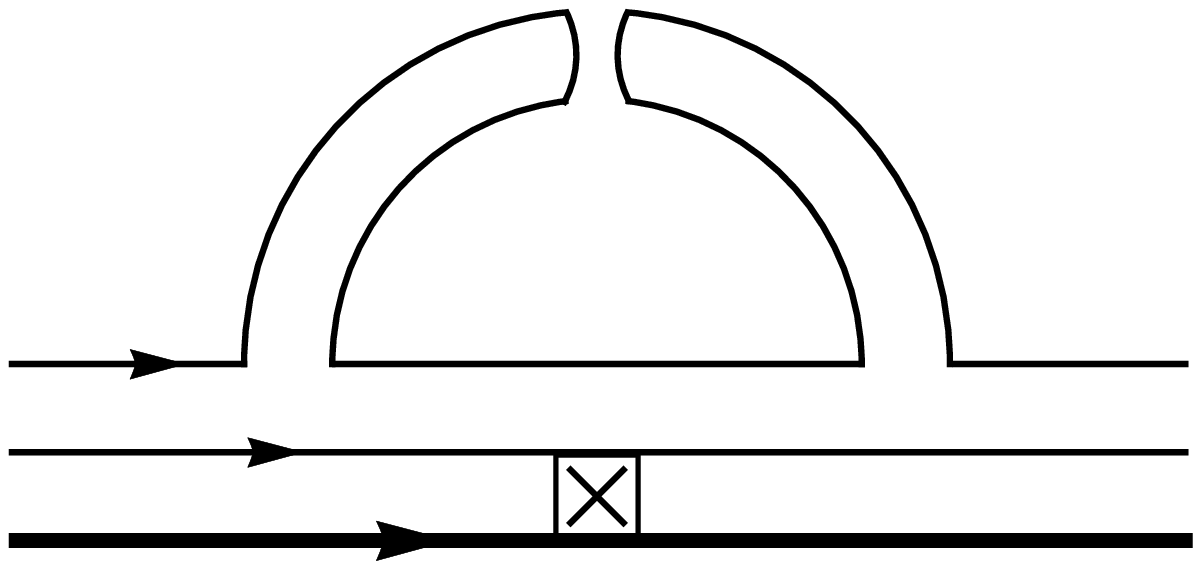}
}
(100,280)*{a};
(1000,280)*{a};
(100,150)*{b};
(1000,150)*{b};
\endxy
}
&
\subfloat[]{
\xy
\xyimport(1100,788){
\includegraphics[width=0.45\textwidth]{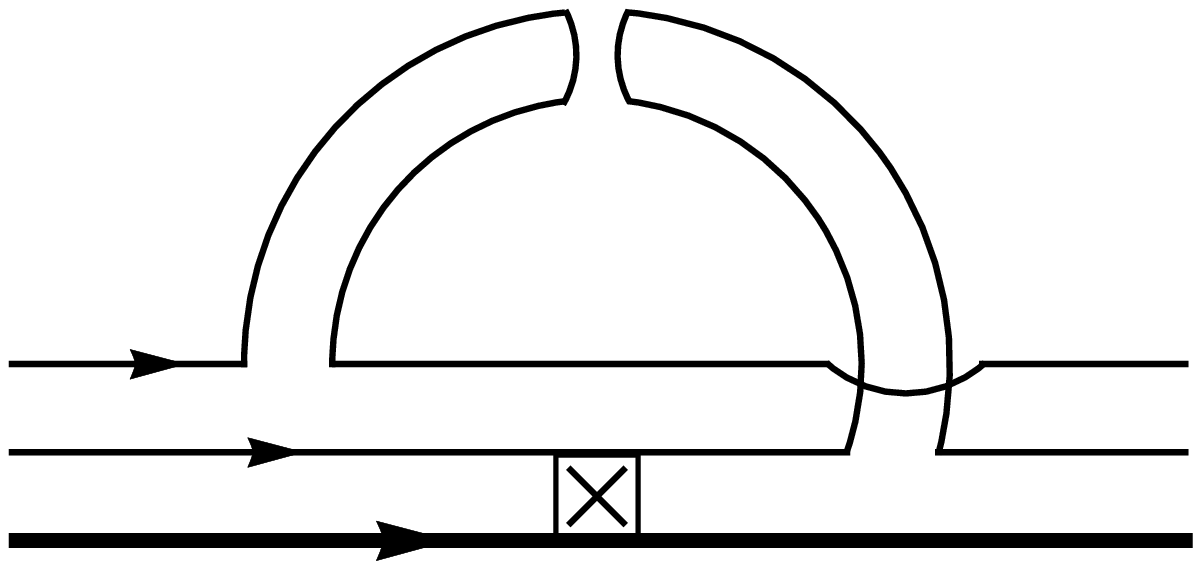}
}
(100,280)*{a};
(1000,280)*{a};
(100,150)*{b};
(1000,150)*{b};
\endxy
}
\\
\subfloat[]{
\xy
\xyimport(1100,788){
\includegraphics[width=0.45\textwidth]{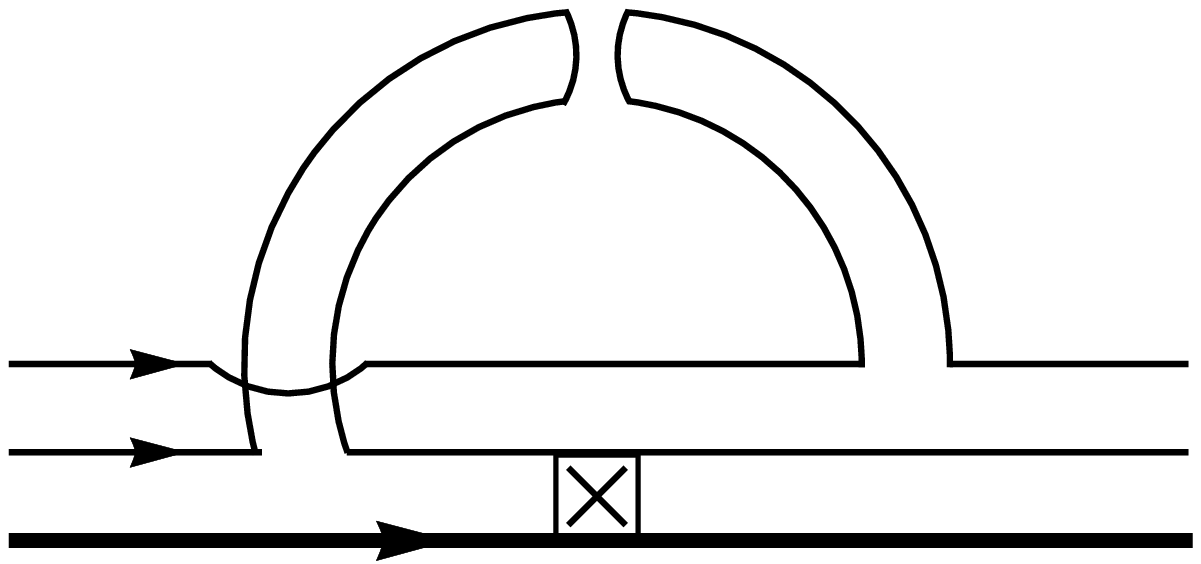}
}
(100,280)*{a};
(1000,280)*{a};
(100,150)*{b};
(1000,150)*{b};
\endxy
}
&
\subfloat[]{
\xy
\xyimport(1100,788){
\includegraphics[width=0.45\textwidth]{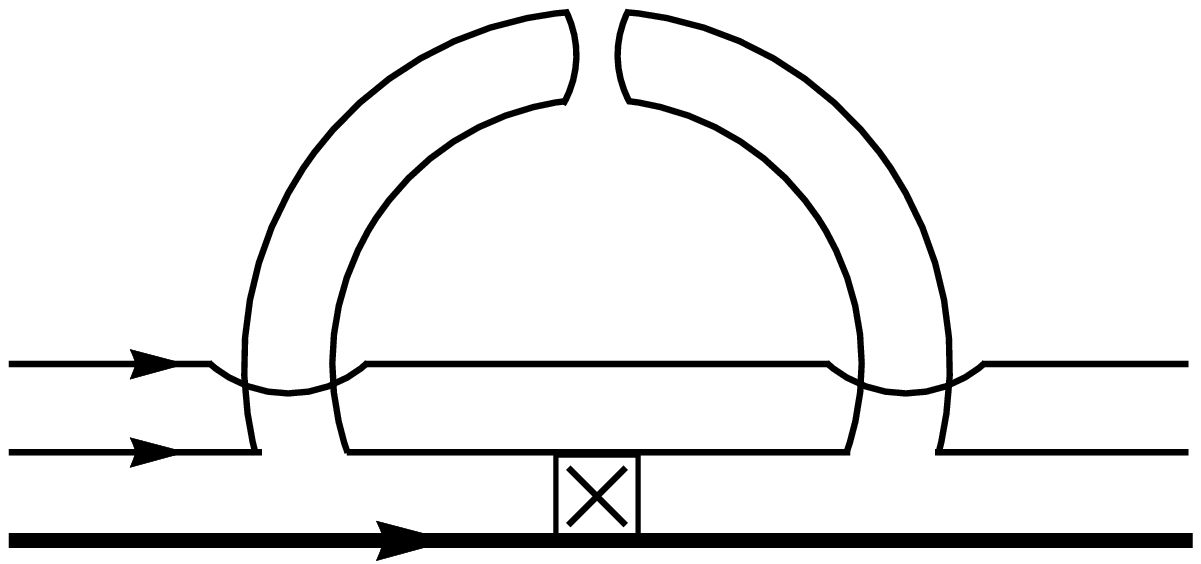}
}
(100,280)*{a};
(1000,280)*{a};
(100,150)*{b};
(1000,150)*{b};
\endxy
}
\end{tabular}
}
\caption{%
\label{fig:sunset_baryon_hairpin}
Sunset diagrams contributing to the $\Delta B=0$ matrix element of single-$b$ baryons involving 
the hairpin structure. 
The solid thick and thin lines denote the anti-$b$ quark and the valence light quark, respectively, 
while the dashed thin line denotes the sea light quark. 
The crossed square is the operator $\tilde{\calO}^\tlam$. 
The diagrams including the internal $S$ and $T$ baryons lead to terms 
multiplied by $\tilde{y}$ and $\tilde{y'}$ in \Eq{sunset_baryon}, 
respectively. 
The diagrams obtained by interchanging the flavor indices $a$ and $b$, 
not shown in this figure, 
have also been taken into account in our results in \Tab{sunset_T_baryon}, 
\Tab{sunset_S_baryon_1} and \Tab{sunset_S_baryon_2}. 
}
\end{figure}

The quark-flavor flow structure for the baryon sunset diagrams 
can be summarized in Figs. \ref{fig:sunset_baryon} and \ref{fig:sunset_baryon_hairpin}. 
Notice that the four-quark operators do not change the spin of a single-$b$ baryon 
as discussed in \Sec{four_quark_ops}, and thus 
there are only three possible combinations of internal baryons, 
$T_{aj}-T_{bj}$, $B_{aj}-B_{bj}$, and $B^*_{aj}-B^*_{bj}$; 
the flavor indices $a$ and $b$ are contracted with $\tlam$, while 
the index $j$ is for the spectator quark. 
For the noneye contraction, both nonhairpin and hairpin diagrams can 
give rise to the connected contribution to the matrix elements; 
the nonhairpin diagram in \Fig{sunset_baryon_2} is the only one 
giving rise to the disconnected contribution. 
The contributions from the sunset diagrams for $T$ and $S$ baryons are summarized as 
\beq
\calQ_T^k&=&\frac{ig_3^2}{f^2}\left[
\sum_\phi (y^{T,k}_{\phi_{ab}}+\lambda^{\tsea}\bar{y}^{T,k}_{\phi_{ab}}) 
\calH(M_{a,b},\Delta^{(B)}+\delta^{(B)}_{a,b})-
\sum_{\phi\phi'}\tilde{y}^{T,k}_{\phi_{aa}\phi'_{bb}}\tilde{\calH}(M_{a,b},\Delta^{(B)})\right],
\nn \\
\calQ_{S_S}^k&=&\frac{ig_2^2}{3f^2}\left[
\sum_\phi (y^{S,k}_{\phi_{ab}}+\lambda^{\tsea}\bar{y}^{S,k}_{\phi_{ab}}) \calH(M_{a,b},\delta^{(B)}_{a,b})-
\sum_{\phi\phi'}\tilde{y}^{S,k}_{\phi_{aa}\phi'_{bb}}\tilde{\calH}(M_{a,b},0)\right],
\\
\calQ_{S_T}^k&=&\frac{ig_3^2}{f^2}\left[
\sum_\phi (y'^{S,k}_{\phi_{ab}}+\lambda^{\tsea}\bar{y}'^{S,k}_{\phi_{ab}}) 
\calH(M_{a,b},-\Delta^{(B)}+\delta^{(B)}_{a,b})-
\sum_{\phi\phi'}\tilde{y'}^{S,k}_{\phi_{aa}\phi'_{bb}}\tilde{\calH}(M_{a,b},-\Delta^{(B)})\right],\nn
\label{eq:sunset_baryon}
\eeq
where the summations are over the Goldstone mesons in SU$(6|3)$ PQ$\chi$PT. 
The coefficients $y, y', \bar{y}, \bar{y}', \tilde{y}$ and $\tilde{y}'$ are presented in 
\Tab{sunset_T_baryon}, \Tab{sunset_S_baryon_1} and \Tab{sunset_S_baryon_2} in \Appendix{coefficients_B_baryon}. 

\subsection{Evaluation of one-loop contributions}
\label{sec:one_loop_evaluation}

In this subsection, we evaluate the one-loop diagrams 
including external $B_d$, $B_s$ mesons, 
and $\Lambda_b$ baryon to provide insight on chiral extrapolations 
of the matrix elements obtained from lattice simulations. 
As seen in Eqs. \ref{eq:mat_elements_B_meson} and \ref{eq:mat_elements_B_baryon}, 
the matrix elements of the singlet four-quark operators are complicated by 
the eye contraction. 
For this reason, we restrict our attention to the matrix elements of the octet operators. 
We carry out the calculations in the QCD limit, 
where the masses of sea quarks are the same with their counterpart valence quarks. 
We fix $f=0.132$ GeV, $\Delta^{(M)}=45$ MeV, $\Delta^{(B)}=200$ MeV, 
$M_{s,s}=691$ MeV, $\tilde{\lambda}_1=0.189~\textrm{GeV}^{-1}$, 
and $\tilde{\lambda}_2=0.377~\textrm{GeV}^{-1}$ \footnote{
The values of $\Delta^{(M)}$ and $\Delta^{(B)}$ 
are consistent with experiment. 
The value of $M_{s,s}$ is determined by the Gell-Mann-Okubo formulas 
using $(M_K)_\textrm{phys}=0.498$ GeV and $(M_\pi)_\textrm{phys}=0.135$ GeV, 
while the values of $\tilde{\lambda}_1$ and $\tilde{\lambda}_2$ 
are determined by \Eq{m_shift_light} using 
$(M_{B_s})_\textrm{phys}-(M_{B_d})_\textrm{phys}=0.087$ GeV 
and $(M_{\Xi^0})_\textrm{phys}-(M_{\Lambda_b^0})_\textrm{phys}=0.173$ GeV, respectively. 
The physical values of hadron masses used in the determination of these  various 
parameters are found in Ref. \cite{Agashe:2014kda}. 
}. 
We also fix the renomalization scale by $\mu=4\pi f$. 
The numerical values of the axial couplings $g_1$ and $g_3$ are taken from the 
recent lattice QCD calculations \cite{Detmold:2011bp,Detmold:2012ge} as follows: $0.398<g_1<0.5$ and $0.58<g_3<0.84$ with 
central values of $g_1=0.449$ and $g_3=0.71$, respectively. 
We also vary the ratios of $P^*$ to $P$ and $S$ to $T$ LECs over the reasonable ranges, 
$|\beta_2/\beta_1|<2$ for $B$ mesons and $|(-\beta'_1+2\beta'_2)/\beta'_3|<2$ for $\Lambda_b$. 

We first consider SU$(3)$ theory in the isospin limit. The typical one-loop contributions 
involve kaons or $\eta$ mesons, 
where their chiral logarithms in the SU$(2)$ chiral limit are given as 
$(M_K^2/\mu^2) \log (M_K^2/\mu^2)\sim 0.21$ or $(M_\eta^2/\mu^2) \log (M_\eta^2/\mu^2)\sim 0.25$. 
Using the pion mass over the typical range of $200~\textrm{MeV} \leq M_\pi \leq 400~\textrm{MeV}$ 
and the standard subtraction scheme in \Eq{MS_subtract}, we find that 
the size of one-loop contributions is comparable with the LO matrix elements. 
Therefore, the convergence of the SU$(3)$ chiral expansion of the matrix elements 
becomes questionable. 
Fortunately, for chiral extrapolations of lattice results 
one only requires knowledge of the pion mass dependence, 
and this can be achieved 
within SU$(3)$ theory by taking the subtraction scheme below inspired by SU$(2)$ 
chiral perturbation theory, 
where the one-loop chiral logarithms vanish in the SU$(2)$ chiral limit \cite{
Tiburzi:2005na,WalkerLoud:2006sa,Detmold:2011rb}.
Notice that the use of different subtraction schemes in one-loop integrals 
results in the finite renormalization of the LO matrix elements, 
but it does not change physical quantities such as the hadron masses. 
We define the nonanalytic functions $I^{\textrm{sub}}(m)$ and $H^{\textrm{sub}}(m,\Delta)$, 
which are relevant to the one-loop calculations, as
\beq
\label{eq:new_sub}
I^{(\textrm{sub})}(m)&=&I(m)-I(m_0),\nn \\
H^{(\textrm{sub})}(m,\Delta)&=&H(m,\Delta)-H(m_0,\Delta_0),
\eeq
where $m_0$ and $\Delta_0$ are the mass of Goldstone mesons and 
the mass difference between external and internal single-$b$ hadrons at zero pion mass, respectively.
We recall that the mass of a Goldstone meson is written as $M^2_{a,b}=B_0(m_a+m_b)$, 
i.e. $M_K^2=(M_{s,s}^2+M_\pi^2)/2$ and $M_\eta^2=(2M_{s,s}^2+M_\pi^2)/3$, 
and the strange quark mass $M_{s,s}$ is fixed. 
The above nonanalytic functions involve logarithms of the masses of Goldstone mesons, 
where for kaons and $\eta$ mesons these logarithms\footnote{
In the SU$(2)$ limit, where the mass of strange 
mesons can be treated on the same order as the chiral symmetry breaking scale, 
such logarithms vanish and we recover the subtraction scheme 
used in the context of SU$(2)$ theory \cite{Detmold:2011rb}.}
are much less sensitive to the pion mass than those for pions.
Note that the subtracted terms are constants and appropriately absorbed by the counterterms. 

\begin{figure}
\includegraphics[width=0.48\textwidth]{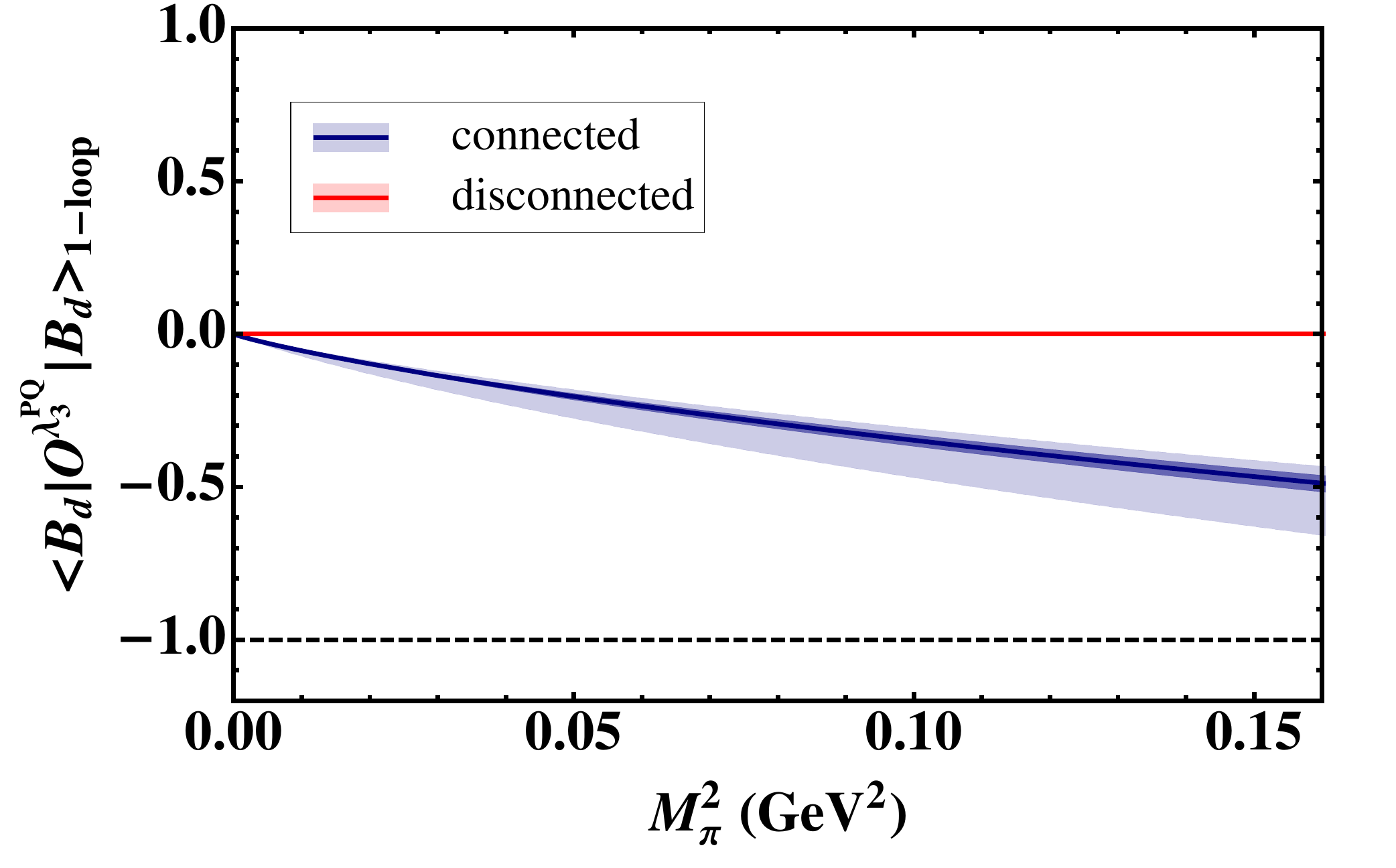}
\includegraphics[width=0.48\textwidth]{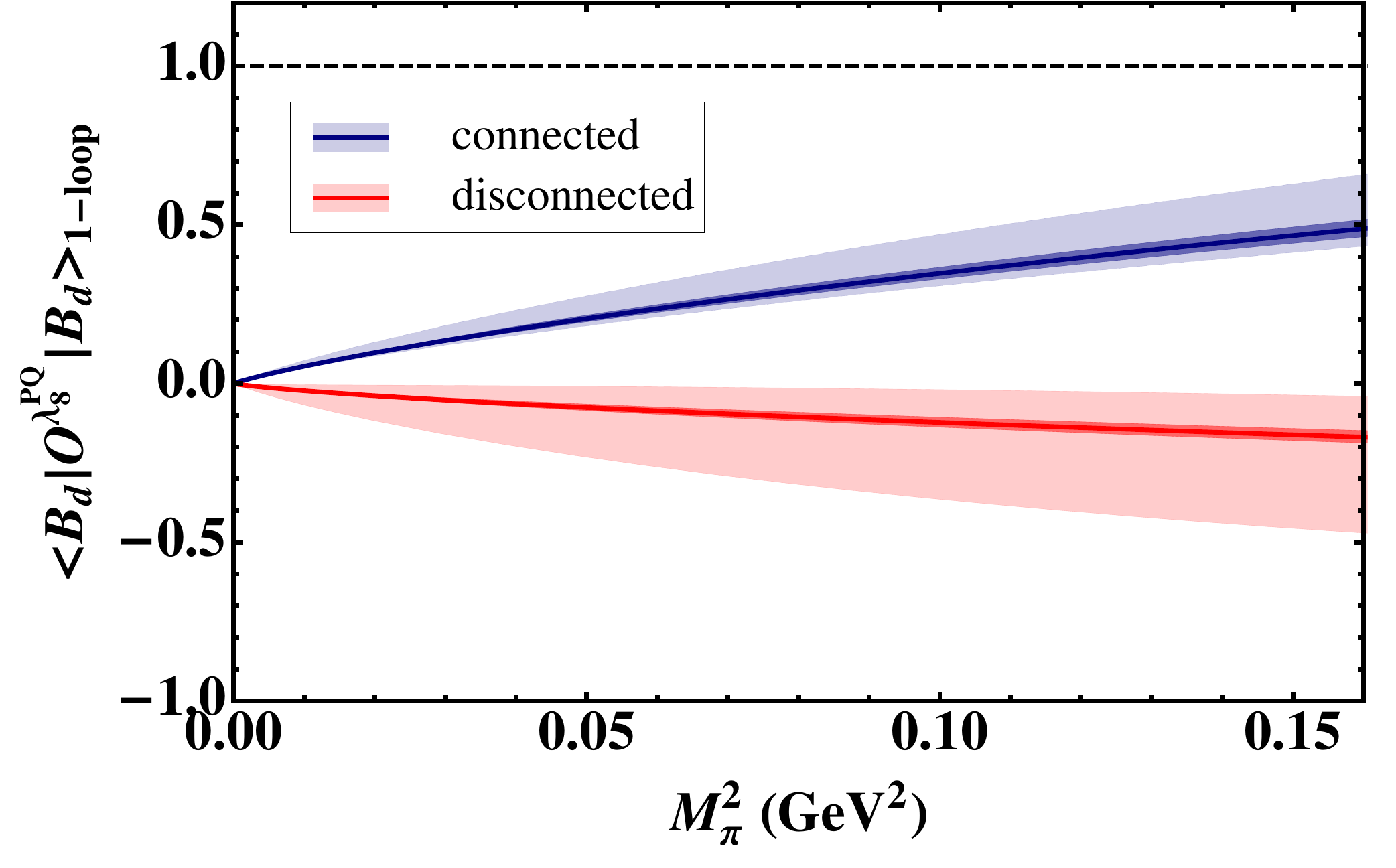}
\includegraphics[width=0.48\textwidth]{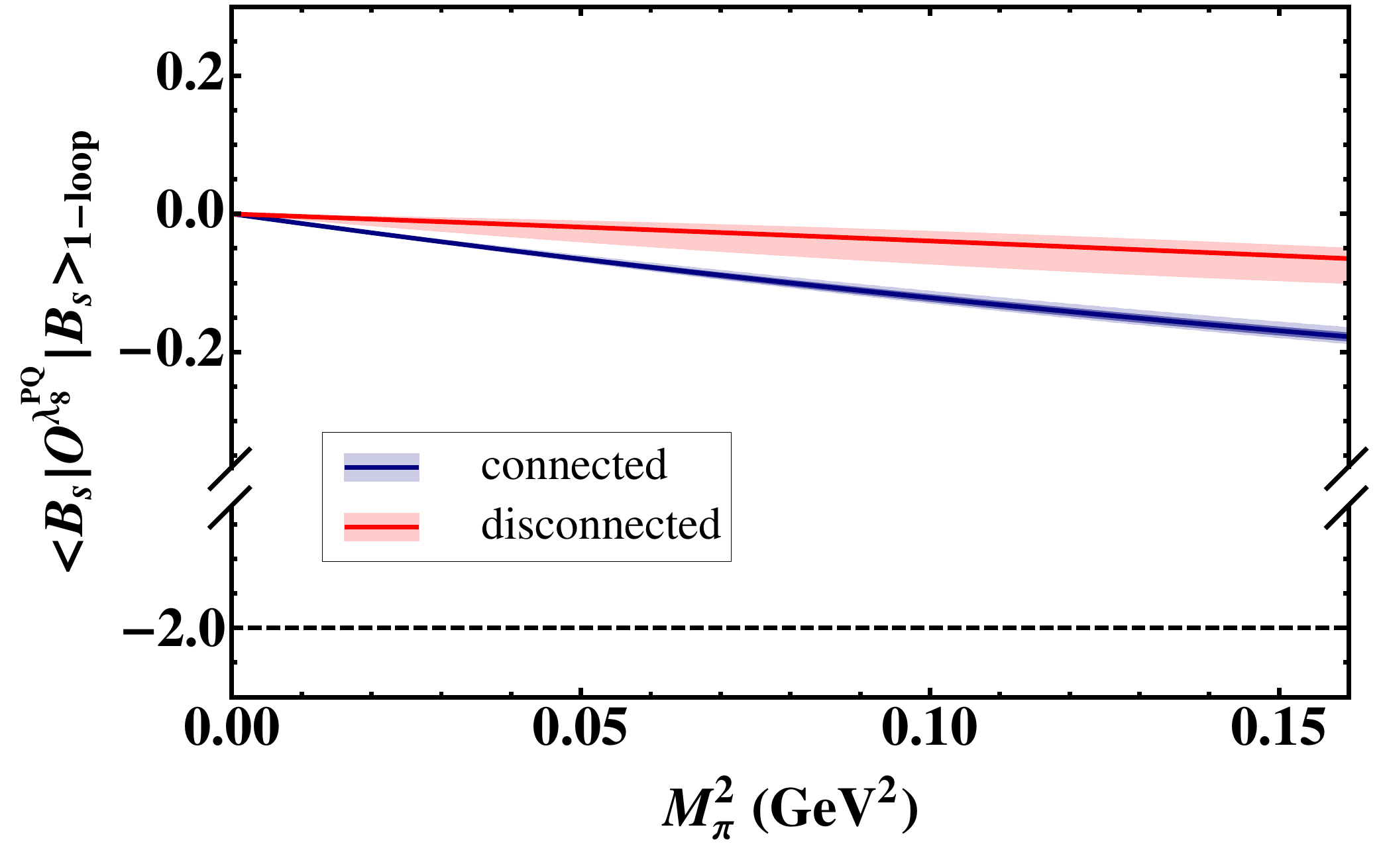}
\caption{%
\label{fig:B_one_loop}
Infinite-volume one-loop contributions to the matrix elements of $\Delta B=0$ four-quark 
operators containing an external $B_d$ for $\lambda^{\textrm{PQ}}_3$ (top left) 
and $\lambda^{\textrm{PQ}}_8$ octets (top right), 
and $B_s$ for $\lambda^{\textrm{PQ}}_8$ octet (bottom). 
For references, black dashed lines correspond to the LO contributions. 
Blue and red colors represent the connected and disconnected one-loop contributions, respectively; 
solid lines are obtained using $g_1=0.449$ and $\beta_{2}/\beta_{1}=1$, 
while inner and outer shaded regions are obtained by varying $g_1$ 
and both $g_1$ and $|\beta_{2}/\beta_{1}|$ over the range given in the text, respectively. 
}
\end{figure}

\begin{figure}
\centerline{
\xy
\xyimport(1,1){
\includegraphics[width=0.48\textwidth]{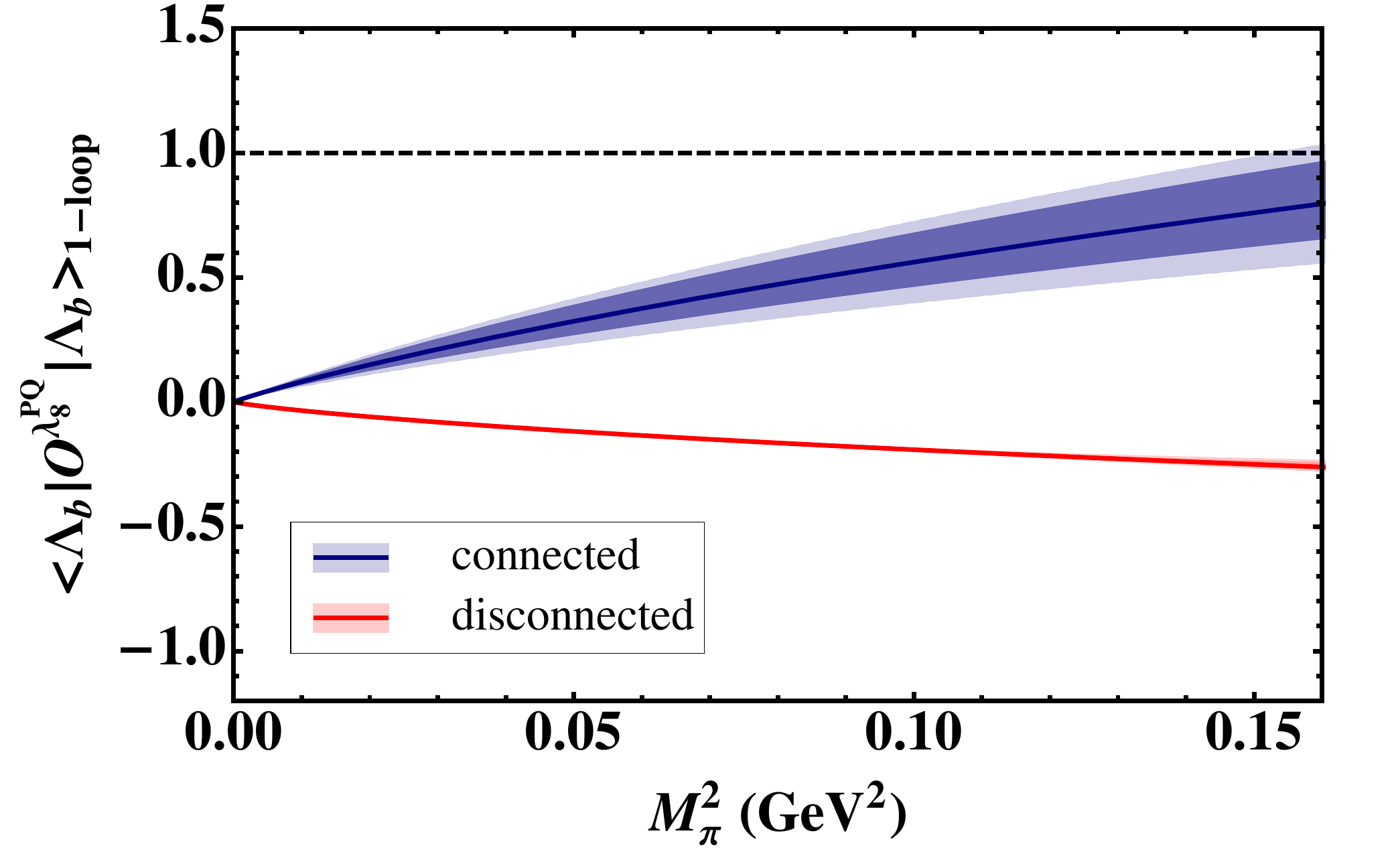}
}
\endxy
}
\caption{%
\label{fig:Lambda_one_loop}
Infinite-volume one-loop contributions to the matrix elements of $\Delta B=0$ four-quark 
operators containing an external $\Lambda_b$ baryon 
for $\lambda^{\textrm{PQ}}_8$ octet. 
For references, black dashed lines correspond to the LO contributions. 
Blue and red colors represent the connected and disconnected one-loop contributions, respectively; 
solid lines are obtained using $g_3=0.71$ and $(-\beta'_1+2\beta'_{2})/\beta'_{3}=1$, 
while inner and outer shaded regions are obtained by varying $g_3$ 
and both $g_3$ and $|(-\beta'_1+2\beta'_{2})/\beta'_{3}|$ over the range given in the text, respectively. 
}
\end{figure}

Our current understanding on the matrix elements are very limited; 
see Ref. \cite{Lenz:2014jha} for the summary of previous determinations. 
In particular, there are only a few quenched lattice simulations 
in Refs. \cite{DiPierro:1998ty,Becirevic:2001fy} for $B$ mesons 
and in Ref. \cite{DiPierro:1999tb} for $\Lambda_b$. 
Instead of using the definite values of the LECs $\beta_1$ and $\beta'_3$, therefore, 
we would rather calculate the NLO one-loop contributions normalized by the LO LECs as follows:
\beq
\langle \mathcal{B}|\calO^{\tlam}| \mathcal{B}\rangle_{\textrm{1-loop}}
=\frac{1}{\beta}\langle \mathcal{B}|\calO^{\tlam}|\mathcal{B}\rangle 
-C^k_{\mathcal{B}},
\eeq
where $\mathcal{B}$ denotes the external single-$b$ hadron, 
while $\beta$ denotes the LECs $\beta_1$ and $\beta'_3$ for $B$ mesons 
and $\Lambda_b$ baryon, respectively. 
Here we use the subtraction scheme defined in \Eq{new_sub} and 
neglect the analytic terms in \Eq{mat_elements_B_meson} 
and \Eq{mat_elements_B_baryon}. 
The results of one-loop calculations are summarized in \Fig{B_one_loop} 
for external $B_d$ and $B_s$, 
and in \Fig{Lambda_one_loop} for $\Lambda_b$. 
In each figure, the black dashed line is 
the LO contribution, 
while the blue and red solid curves are the connected and 
disconnected one-loop contributions, respectively; 
we use the central values of $g_1$ and $g_3$ with the ratios of LECs being unity, 
$\beta_2/\beta_1=(-\beta'_1+2\beta'_2)/\beta'_3=1$. 
The variations of the axial couplings and of both the axial couplings and the ratios of LECs 
are represented by the inner dark-shaded and outer light-shaded regions, respectively. 
It is clear from these figures that 
the sunset diagrams provide an important part of the disconnected one-loop contributions 
to the matrix elements, and thus invoke relatively large uncertainties 
due to the unknown LECs for $B^*$ mesons and $S$-type baryons. 
In addition, the matrix elements containing external $B_s$ 
have relatively small pion-mass dependence compared to those containing $B_d$ 
because of the absence of pion loops, 
where their logarithmic behavior almost fades away. 

For idealized values of LECs, namely $\beta_2=\beta_1$ and $\beta'_3=-\beta'_1+2\beta'_2$, 
the disconnected diagrams provide a sizeable contribution to the matrix elements 
for the $\lambda^{\textrm{PQ}}_8$ octet 
because of large SU$(3)$ breaking effects in the light-quark sector. 
In the case of the $\lambda^{\textrm{PQ}}_3$ octet, the matrix elements 
containing $B_s$ and $\Lambda_b$ do not exist to all orders in the chiral expansion 
provided isospin symmetry is exact. 
As seen in \Fig{B_one_loop}, on the other hand, the matrix element 
containing $B_d$ receives connected one-loop contributions 
which are exactly opposite to those for the $\lambda^{\textrm{PQ}}_8$ octet, 
but it receives no disconnected contributions. 
Since the eye-contractions also vanish, 
one can in principle determine this matrix element very accurately 
from the calculation of connected diagrams which are accessible by current lattice techniques. 
Moreover, its precise value will play a crucial role in the investigation of 
the lifetime ratio $\tau(B^+)/\tau(B_d)$ \cite{Lenz:2014jha}. 

\begin{figure}
\includegraphics[width=0.48\textwidth]{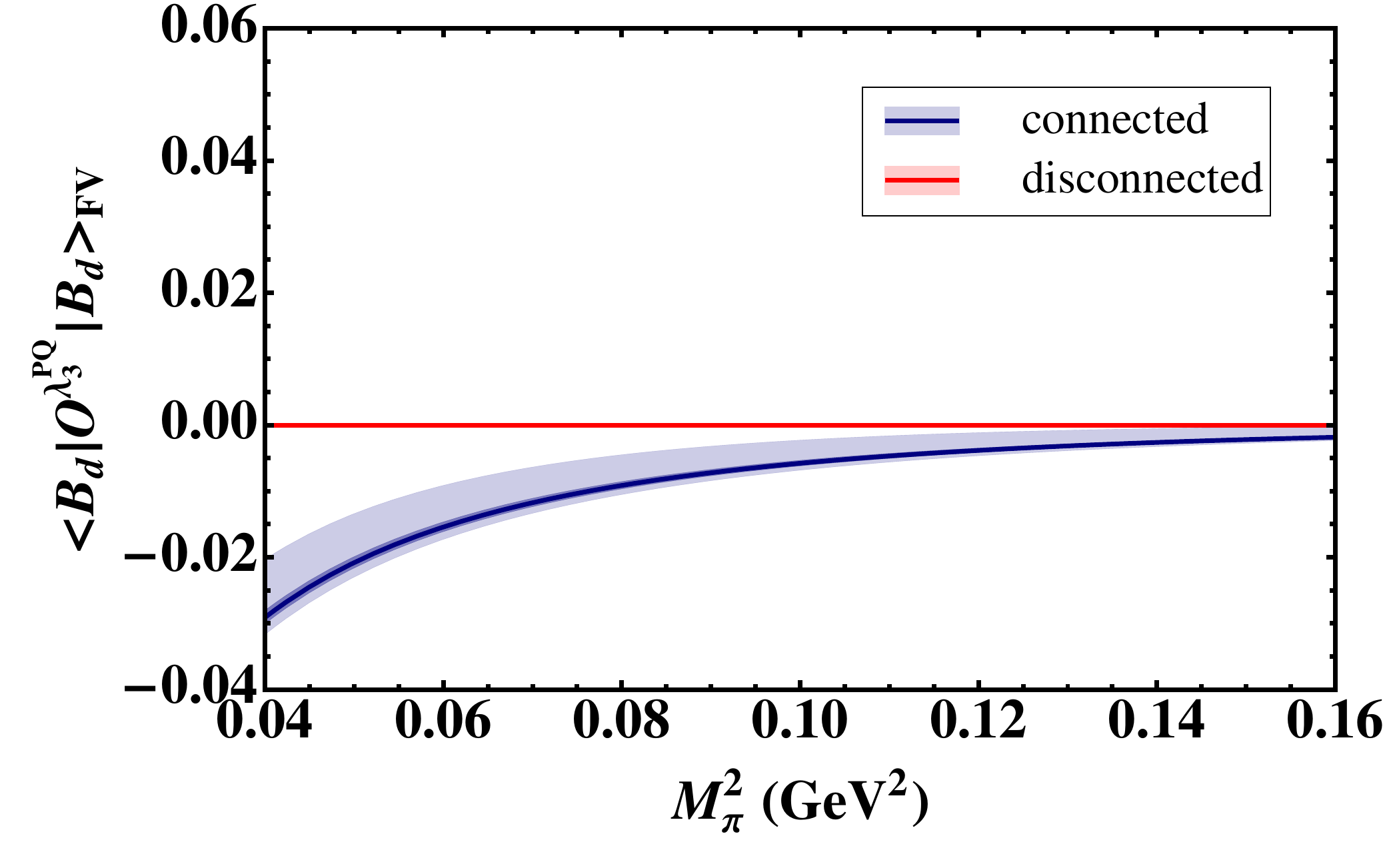}
\includegraphics[width=0.48\textwidth]{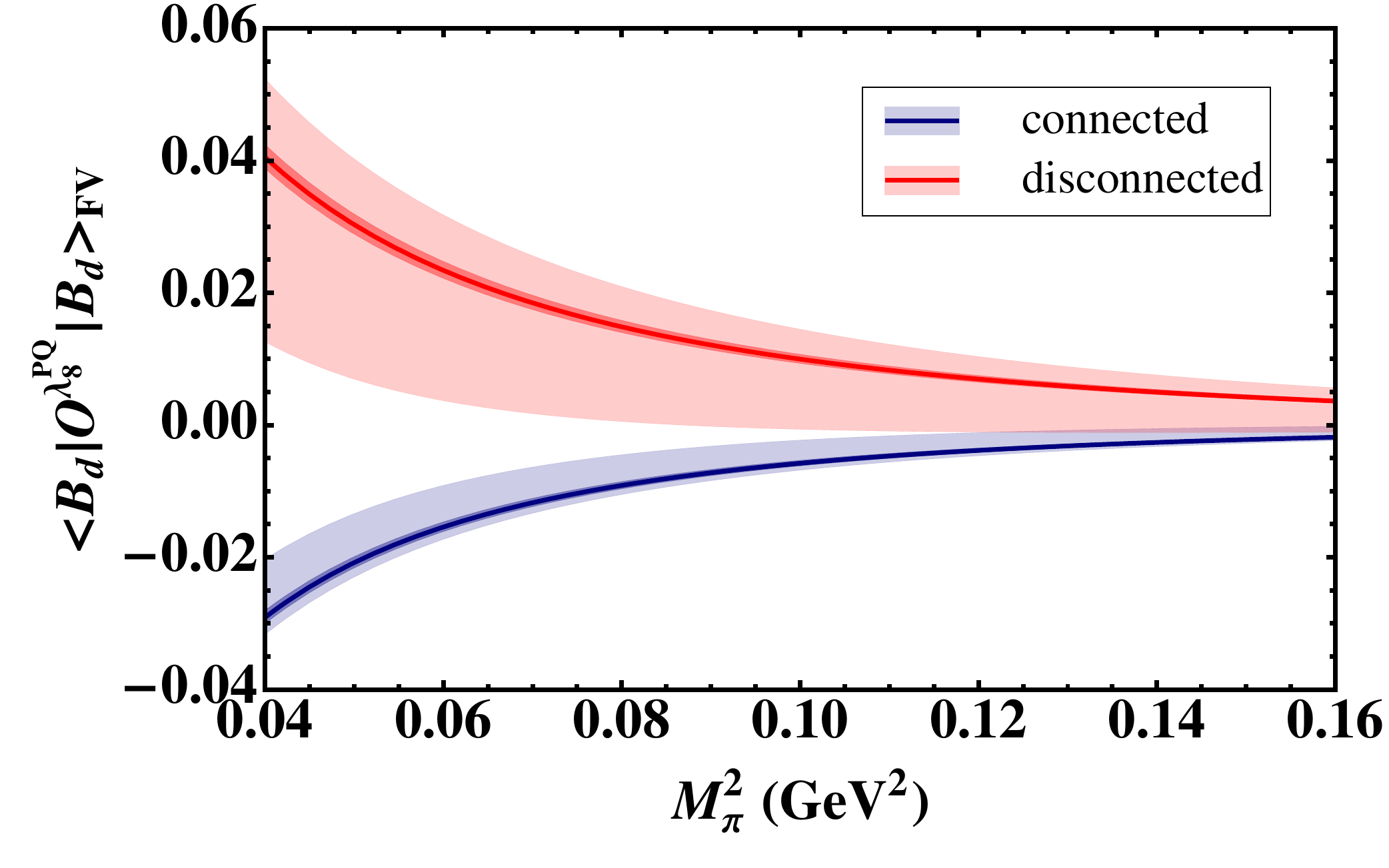}
\includegraphics[width=0.48\textwidth]{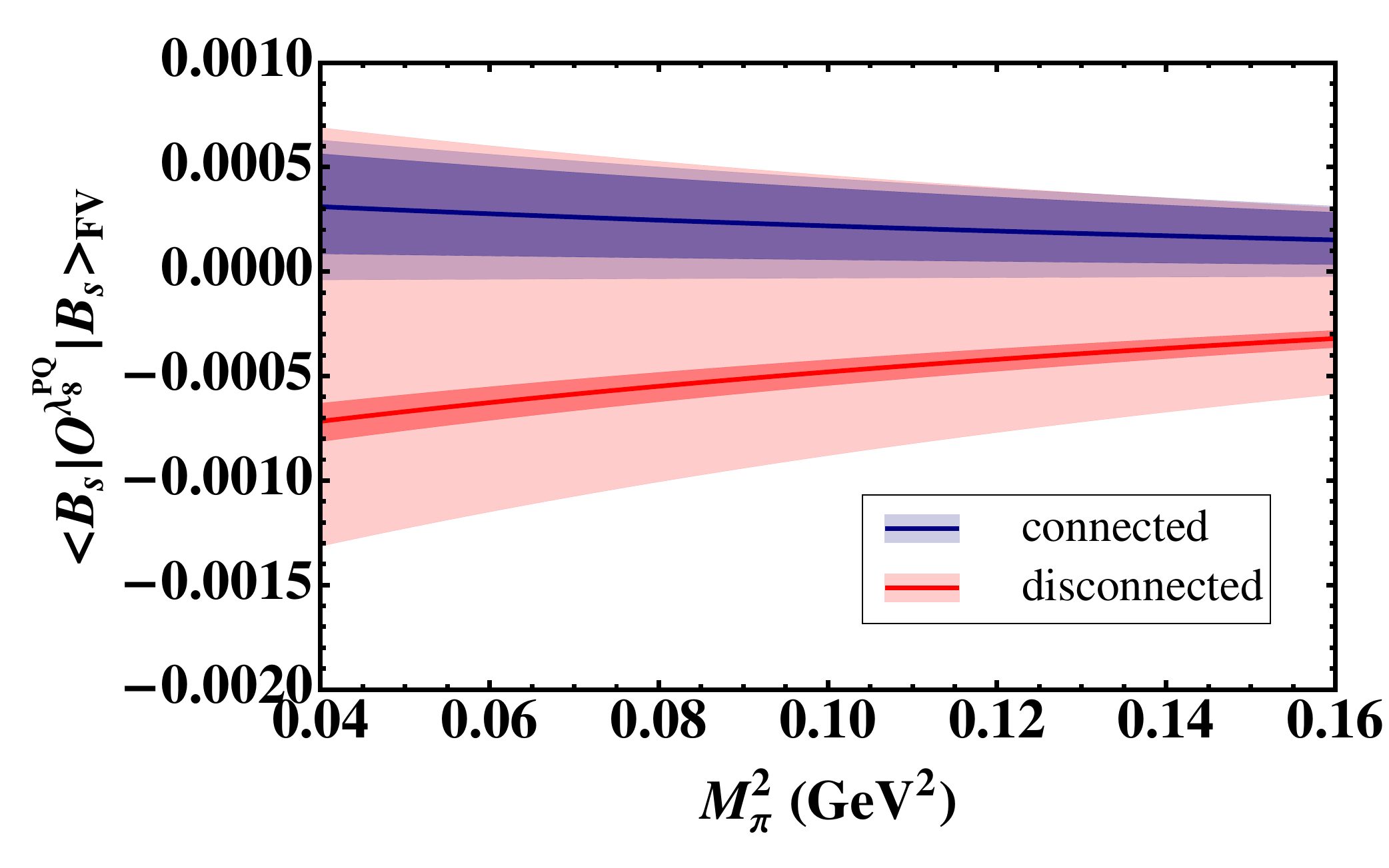}
\caption{%
\label{fig:B_FV_effects}
Finite volume effects in the matrix elements of $\Delta B=0$ four-quark operators 
containing external $B_d$, 
for $\lambda^{\textrm{PQ}}_3$ (top left) and $\lambda^{\textrm{PQ}}_8$ octets (top right), 
and $B_s$ for $\lambda^{\textrm{PQ}}_8$ octet (bottom) 
over the range of $m_\pi L$ from $2.5$ to $5$ in an $L=2.5$fm lattice. 
The details are the same as those in \Fig{B_one_loop}. 
}
\end{figure}

\begin{figure}
\includegraphics[width=0.48\textwidth]{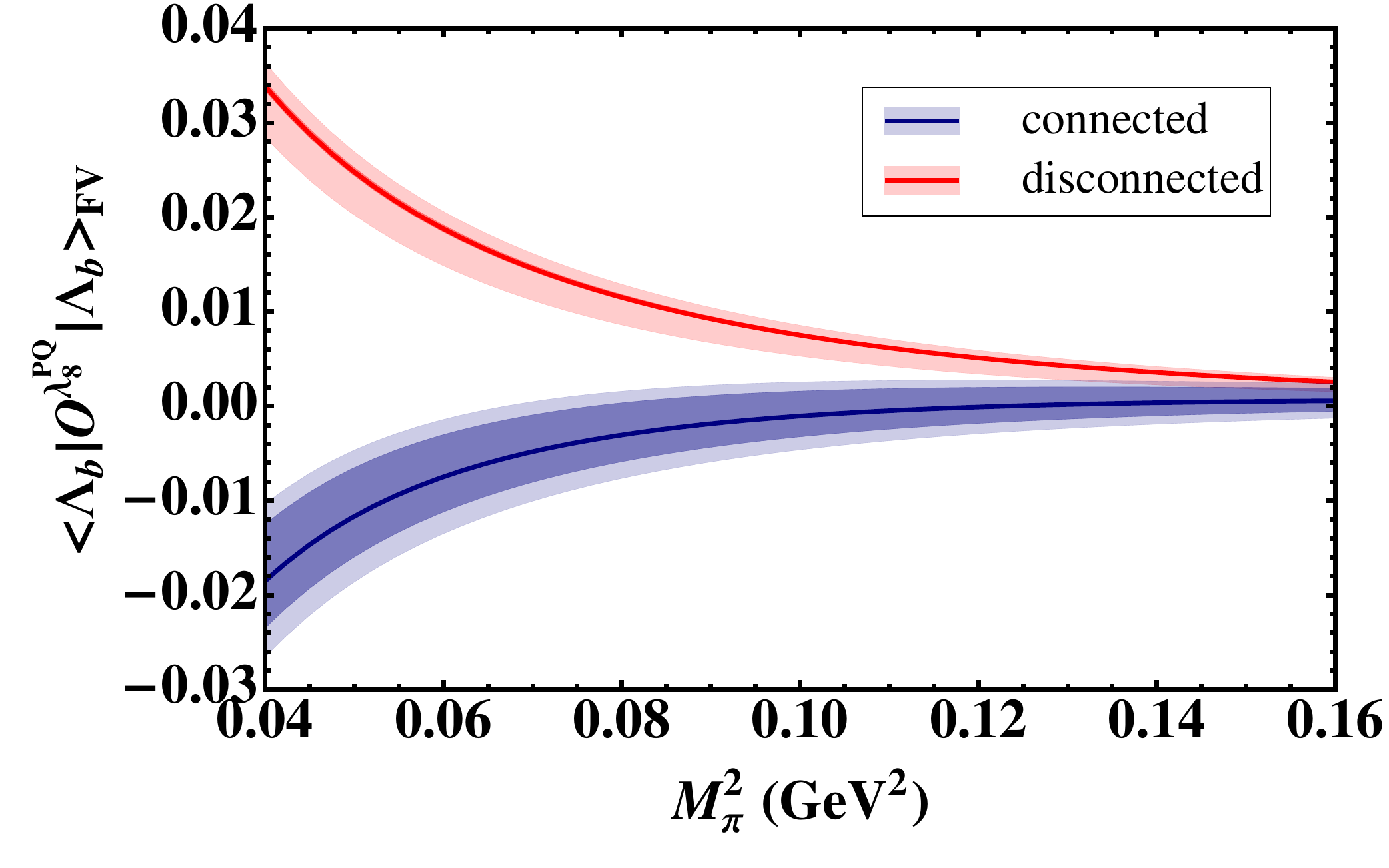}
\caption{%
\label{fig:Lambda_b_FV_effects}
Finite volume effects in the matrix elements of $\Delta B=0$ four-quark operators 
containing an external $\Lambda_b$ baryon 
for $\lambda^{\textrm{PQ}}_8$ octet 
over the range of $m_\pi L$ from $2.5$ to $5$ in an $L=2.5$fm lattice. 
The details are the same as those in \Fig{Lambda_one_loop}. 
}
\end{figure}

Now, we explore the finite volume effects in 
$\Delta B=0$ matrix elements for external $B_d$, $B_s$ mesons, and $\Lambda_b$ baryon. 
To this end, we consider the change in the finite volume matrix elements relative to 
the infinite ones normalized by their three-level values, 
\beq
\langle \calB|\calO^\tlam|\calB\rangle_{\textrm{FV}}
=\frac{\langle \calB|\calO^\tlam|\calB\rangle(L)
-\langle\calB|\calO^\tlam|\calB\rangle(\infty)}
{\langle\calB|\calO^\tlam|\calB\rangle_{\textrm{tree}}},
\label{eq:FV_matrix_elements}
\eeq
where $\calB$ denotes the external single-$b$ hadron. 
With a lattice volume fixed by $L=2.5$ fm we present the results for $B_d$, $B_s$ 
in \Fig{B_FV_effects}, and for $\Lambda_b$ in \Fig{Lambda_b_FV_effects}, 
where the blue and red solid curves are the connected and 
disconnected one-loop contributions; 
we use the central values of $g_1$ and $g_3$ with the ratios of LECs being unity, 
$\beta_2/\beta_1=(-\beta'_1+2\beta'_2)/\beta'_3=1$. 
The variations of the axial couplings and of both the axial coupling and the ratios of LECs 
are represented by the inner dark-shaded and outer light-shaded regions, respectively. 
In the case of $B_d$, 
finite volume effects in the connected diagrams for $\lambda^{\textrm{PQ}}_3$ are exactly the same as 
those for the $\lambda^{\textrm{PQ}}_8$ octet, 
while the effects in the disconnected diagrams vanish. 
In the cases of $B_s$ and $\Lambda_b$, 
the change in \Eq{FV_matrix_elements} for $\lambda^{\textrm{PQ}}_3$ is ill defined 
because the corresponding tree-level matrix elements do not exist.
Finite volume effects for $B_s$ are considerably suppressed 
because pion loops do not contribute to these matrix elements in the QCD limit, 
while the sizes of finite volume effects for $\Lambda_b$ are similar to those 
for $B_d$. As seen in these figures, 
the dominant uncertainty in our calculations, 
especially for the disconnected contributions, 
arises from the variation of the ratio of unknown LECs. 

\section{Conclusion}
\label{sec:conclusion}

We have determined the matrix elements of $\Delta B=0$ four-quark operators 
involving single-$b$ hadron external states 
relevant for the 
calculation of $V_{ub}$ and lifetimes of single-$b$ hadrons 
at the next-to-leading order in heavy hadron chiral perturbation theory. 
We performed these calculations in SU$(6|3)$ partially quenched theories 
including finite volume effects in the isospin limit, 
where connected and disconnected contributions are analyzed in a natural manner. 
At the tree level, the disconnected contributions called the eye contractions 
vanish for the octet operators due to the light-flavor symmetry, but 
they do not for the singlet operator. 
At the next-to-leading order in chiral expansion for the exemplified cases of 
external $B_d$, $B_s$, and $\Lambda_b$, 
we find that 
for the $\lambda^{\textrm{PQ}}_8$ octet 
the disconnected contributions are suppressed but still 
sizeable due to the large light-flavor symmetry breaking. 
In the case of the $\lambda^{\textrm{PQ}}_3$ octet with 
an external $B_d$ meson, on the other hand, 
we find that the matrix elements 
receive contributions solely from connected diagrams. 
This result suggests that the precise calculation of these particular matrix elements, 
which have phenomenological impact on the determination of the 
lifetime ratio $\tau(B^+)/\tau(B_d)$, 
is possible using current lattice techniques. 

Additionally, our results are necessary to extrapolate the QCD and PQQCD lattice data 
of these matrix elements 
from the unphysical light-quark masses used in the finite volume lattice simulations 
to their physical values. 
We find that such chiral extrapolations are complicated by more than two 
unrelated low-energy constants appearing in the leading-order 
matrix elements in Eqs. \ref{eq:O_hmchipt} and \ref{eq:O_hhchipt}.
Finite volume effects are at the few-percent level for a lattice volume of $L=2.5$ fm, 
and should be taken into account for the extrapolations 
as they become important in high-precision calculations.

\acknowledgments
The author would like to acknowledge William Detmold and C.-J. David Lin 
for their contributions to the early stage of this work, 
and for helpful discussions. 
The author also thanks Brian C. Tiburzi for many fruitful discussions and comments. 
This work is supported by the U.S. National Science Foundation, 
under Grant No. PHY12-05778. 

\bibliography{deltaB0}

\appendix

\section{Integrals and sums}
\label{sec:integrals_and_sums}

The ultraviolet divergences in various loop integrals are regularized by dimensional regularization 
and have the term
\beq
\bar{\lambda}=\frac{2}{4-d}-\gamma_E+\log (4\pi) +1,
\label{eq:MS_subtract}
\eeq
where $d$ is the number of space-time dimension. 
This is a commonly used scheme in $\chi$PT calculations \cite{Gasser:1984gg}, 
which is different from the $\overline{\textrm{MS}}$ scheme by 
the constant ``$1$'' on the right-hand side of the above equation. 
The requisite subtracted integrals appearing in the one-loop calculations 
in the framework of PQ$\chi$PT can all be obtained from the following integrals, 
\beq
I(m)&\equiv& \mu^{4-d}\int \frac{d^dk}{(2\pi)^d}\frac{1}{k^2-m^2+i\epsilon}-\frac{im^2}{16\pi^2}\bar{\lambda}\nn \\
&=&-\frac{im^2}{16\pi^2}\log \left(\frac{m^2}{\mu^2}\right),
\label{eq:integral_I}
\eeq
\beq
F(m,\Delta)&\equiv&\left(g^{\rho \nu}-v^\rho v^\nu\right)\left[\frac{\mu^{4-d}}{(d-1)}\int\frac{d^dk}{(2\pi)^d}\frac{k_\rho k_\nu}{(k^2-m^2+i\epsilon)(v\cdot k-\Delta+i\epsilon)}\right.\nn \\
&&~~~~~~~~~~~~~~~~~~~~~~~~~~~~~~~~~~~~~~~~~~~~~~~~~~~~~~~~~~~~~~~~~~~~~~~\left.-\frac{ig_{\rho \nu}}{16\pi^2}\bar{\lambda}\left(\frac{2\Delta^2}{3}-m^2\right)\Delta\right]\nn \\
&=&\frac{i}{16\pi^2}\left[\log \left(\frac{m^2}{\mu^2}\right)\left(m^2-\frac{2}{3}\Delta^2\right)\Delta
+\left(\frac{10}{9}\Delta^2-\frac{4}{3}m^2\right)\Delta+\frac{2(\Delta^2-m^2)}{3}mR\left(\frac{\Delta}{m}\right)\right],\nn \\
\label{eq:integral_F}
\eeq
where $\mu$ is the renormalization scale and 
with
\beq
R(x)\equiv \sqrt{x^2-1}\log \left(\frac{x-\sqrt{x^2-1+i\epsilon}}{x+\sqrt{x^2-1+i\epsilon}}\right).
\eeq

To study finite volume effects in the limit $mL\gg1$, 
we consider a cubic spatial box of $V=L^3$ with periodic boundary conditions. 
The three-momenta are quantized as $\vec{k}=\left(2\pi/L\right)\vec{e}$, 
where $e_1,e_2,e_3=-L/2+1,\cdots,L/2$. 
After subtracting the ultraviolet divergence like as in \Eq{integral_I} and \Eq{integral_F}, 
we instead obtain the sums
\beq
\calI(m)&\equiv& \frac{1}{L^3}\sum_{\vec{k}} \int \frac{dk_0}{2\pi}\frac{1}{k^2-m^2+i\epsilon}\nn \\
&=&I(m)+I_{FV}(m),
\eeq
\beq
\calF(m,\Delta) &\equiv& \frac{(g^{\rho \nu}-v^\rho v^\nu)}{(d-1)}\frac{1}{L^3}\sum_{\vec{k}}\int\frac{dk_0}{2\pi}\frac{k_\rho k_\nu}{(k^2-m^2+i\epsilon)(v\cdot k-\Delta+i\epsilon)}\nn \\
&=&F(m,\Delta)+F_{FV}(m,\Delta).
\eeq
The finite-volume pieces are given by \cite{Arndt:2004bg,Detmold:2006gh,Detmold:2011rb}
\beq
I_{FV}(m)&=&\frac{-i}{4\pi^2}m\sum_{\vec{n}\neq\vec{0}}\frac{1}{nL}K_1(nmL)\nn \\
&\overset{mL\gg1}{\longrightarrow}&\frac{-i}{4\pi^2}\sum_{\vec{n}\neq\vec{0}}\sqrt{\frac{m\pi}{2nL}}
\left(\frac{1}{nL}\right)e^{-nmL}\times\left\{1+\frac{3}{8nmL}-\frac{15}{128(nmL)^2}
+\calO\left(\left[\frac{1}{nmL}\right]^3\right)\right\},\nn \\
F_{FV}(m,\Delta)&=&\frac{i}{12\pi^2}\sum_{\vec{n}\neq\vec{0}}\frac{1}{nL}
\int_0^\infty d|\vec{k}| \frac{|\vec{k}|\sin(n|\vec{k}|L)}{\sqrt{|\vec{k}|^2+m^2}+\Delta}
\left(\Delta+\frac{m^2}{\sqrt{|\vec{k}|^2+m^2}}\right),\nn \\
&\overset{mL\gg1}{\longrightarrow}&\frac{i}{24\pi}m^2\sum_{\vec{n}\neq\vec{0}}\frac{e^{-nmL}}{nL}\calA
\eeq
where $\vec{n}=(n_1,n_2,n_3)$ with $n_i\in \mathbb{Z}$, $n\equiv |\vec{n}|$, and
\beq
\calA&=&e^{(z^2)}[1-\textrm{Erf}(z)]+\left(\frac{1}{nmL}\right)\left[\frac{1}{\sqrt{\pi}}
\left(\frac{9z}{4}-\frac{z^3}{2}\right)+\left(\frac{z^4}{2}-2z^2\right)e^{(z^2)}[1-\textrm{Erf}(z)]\right]\nn\\
&&+\left(\frac{1}{nmL}\right)^2\left[\frac{1}{\sqrt{\pi}}\left(\frac{39z}{64}-\frac{11z^3}{32}
+\frac{9z^5}{16}-\frac{z^7}{8}\right)-\left(\frac{z^6}{2}-\frac{z^8}{8}\right)e^{(z^2)}
[1-\textrm{Erf}(z)]\right]+\calO \left(\left[\frac{1}{nmL}\right]^3\right),\nn
\eeq
with
\beq
z\equiv \left(\frac{\Delta}{m}\right)\sqrt{\frac{nmL}{2}}.\nn
\eeq
The functions appearing in the one-loop results are
\beq
\calH(m,\Delta) \equiv \frac{\partial \calF(m,\Delta)}{\partial \Delta},
\eeq
and 
\beq
\tilde{\calH}(M_{a,b},\Delta)
&=&\delta_{a,b}\lbrace A_{a,b}\calH(M_{a,b},\Delta)
+(1-A_{a,b})\calH(M_X,\Delta)+C_{a,b}\calH_{\eta'}(M_{a,b},\Delta)
\rbrace
\nn\\
&&+(1-\delta_{a,b})\lbrace
D^{(a)}_{a,b}\calH(M_{a,a},\Delta)
+D^{(b)}_{a,b}\calH(M_{b,b},\Delta)+D^{(X)}_{a,b}\calH(M_X,\Delta)\rbrace,
\eeq
where the function $\calH_{\eta'}$ from the integral of hairpin diagrams is 
\beq
\calH_{\eta'}(m,\Delta) \equiv \frac{\partial \calH(m,\Delta)}{\partial m^2}.
\eeq
The coefficients in the function $\tilde{\calH}$ are defined by
\beq
&&A_{u,u}=\frac{2\left(\delta^2_{VS}-M^2_\pi+M^2_X\right)\delta^2_{VS}}{(M^2_\pi-M^2_X)^2}+\frac{3}{2},~~~
A_{s,s}=\frac{3\left(8\delta^4_{VSs}+\left(2\delta^2_{VS}-M^2_\pi+M^2_{s,s}\right)^2\right)}
{\left(2\delta^2_{VS}+4\delta^2_{VSs}-M^2_\pi+M^2_{s,s}\right)^2},\nn \\
&&C_{u,u}=3\delta^2_{VS}-\frac{2\delta^4_{VS}}{M^2_\pi-M^2_X},~~~
C_{s,s}=\frac{6\delta^2_{VSs}\left(2\delta^2_{VS}-M^2_\pi+M^2_{s,s}\right)}{2\delta^2_{VS}+4\delta^2_{VSs}-M^2_\pi+M^2_{s,s}},\nn \\
&&D^{(u)}_{u,s}=\frac{2\delta^2_{VS} \left(M_\pi^2-M_{s,s}^2+2\delta^2_{VSs}\right)}
{(M^2_\pi-M^2_{s,s})(M^2_\pi-M^2_X)},~~~
D^{(s)}_{u,s}=\frac{2\delta^2_{VSs} \left(M_\pi^2-M_{s,s}^2-2\delta^2_{VS}\right)}
{(M^2_\pi-M^2_{s,s})(M^2_{s,s}-M^2_X)},\nn \\
&&D^{(X)}_{u,s}=\frac{(M^2_\pi-M^2_X-2\delta^2_{VS})(M_{s,s}^2-M_X^2-2\delta^2_{VSs})}
{(M^2_\pi-M^2_X)(M^2_{s,s}-M^2_X)},
\eeq
where $M_{a,b}^2=B_0(m_a+m_b)$, $\delta^2_{VS}=M^2_{u,u}-M^2_{u,j}$, 
$\delta^2_{VSs}=M^2_{s,s}-M^2_{s,r}$, $M_\pi=M_{u,u}$, 
and $M^2_X=\frac{1}{3}\left(M^2_{u,u}+2M^2_{s,s}-2\delta^2_{VS}-4\delta^2_{VSs}\right)$. 
Notice that we take the isospin limit as in \Eq{quark_mass_matrix}. 
In the QCD limit (setting valence and sea quark masses to be identical), we have
\beq
&&A^{QCD}_{u,u}=\frac{3}{2},~~~A^{QCD}_{s,s}=3,~~~C^{QCD}_{u,u}=0,~~~C^{QCD}_{s,s}=0,\nn \\
&&D^{(u)QCD}_{u,s}=0,~~~D^{(s)QCD}_{u,s}=0,~~~D^{(X)QCD}_{u,s}=1.
\eeq

\section{coefficients for chiral one-loop contributions in $B$ meson}
\label{sec:coefficients_B_meson}

In this appendix, we present the coefficients in \Eq{tadpole_meson} 
and \Eq{sunset_meson} corresponding to the tadpole- and sunset-type 
diagrams. These coefficients are summarized in \Tab{tadpole_meson} and \Tab{sunset_meson}. 

\begin{table}
\caption{%
\label{tab:tadpole_meson}
Coefficients $x^{B,k}_{\phi_{ab}}$ and $\bar{x}^{B,k}_{\phi_{ab}}$ in \Eq{tadpole_meson} 
in the isospin limit. 
In the case of $B_d$, the coefficients for $k=0,8$ and for $k=3$ 
are identical and opposite to those of $B_u$, respectively. 
The coefficients for $B_s$ with $k=3$ are all zeros. 
}
\begin{center}
\begin{tabular}{c|c|cccc|cccc}
\multicolumn{2}{c|}{}&\multicolumn{4}{c|}{$x^{B,k}_\phi$}
&\multicolumn{4}{c}{$\bar{x}^{B,k}_{\phi}$}\\
\cline{3-10}
\multicolumn{2}{c|}{}&~~~$uj$~~~&~~~$ur$~~~&~~~$sj$~~~&~~~$sr$~~~
&~~~$uj$~~~&~~~$ur$~~~&~~~$sj$~~~&~~~$sr$~~~\\
\hline
\multirow{2}{*}{~~~$k=0$~~~} & $~~~B_u~~~$ & $-2$ & $-1$ & 0 & 0 &
$2$ & $1$ & 0 & 0 \\
& $B_s$ & 0 & 0 & $-2$ & $-1$ & 
0 & 0 & $2$ & $1$ \\
\hline
$k=3$ & $B_u$ & $-2$ & $-1$ & 0 & 0 & 
0 & 0 & 0 & 0 \\
\hline 
\multirow{2}{*}{$k=8$} & $B_u$ & $-2$ & $-1$ & 0 & 0 & 
$2$ & $-2$ & 0 & 0 \\
& $B_s$ & 0 & 0 & $4$ & $2$ & 
0 & 0 & $2$ & $-2$ \\
\hline\hline 
\end{tabular}
\end{center}
\end{table}
\begin{table}
\caption{%
\label{tab:sunset_meson}
Coefficients $\bar{y}^{B,k}_{\phi_{ab}}$ and $\tilde{y}^{B,k}_{\phi_{aa}\phi'_{bb}}$ 
in \Eq{sunset_meson} in the isospin limit. 
In the case of $B_d$, the coefficients for $k=0,8$ and for $k=3$ 
are identical and opposite to those of $B_u$, respectively. 
The coefficients for $B_s$ with $k=3$ are all zeros.
}
\begin{center}
\begin{tabular}{c|c|cccc|ccc}
\multicolumn{2}{c|}{}&\multicolumn{4}{c|}{$\bar{y}^{B,k}_{\phi}$}
&\multicolumn{3}{c}{$\tilde{y}^{B,k}_{\phi\phi'}$}\\
\cline{3-9}
\multicolumn{2}{c|}{}&~~~$uj$~~~&~~~$ur$~~~&~~~$sj$~~~&~~~$sr$~~~
&~~~$\eta_u\eta_u$~~~&~~~$\eta_u\eta_s$~~~&~~~$\eta_s\eta_s$~~~\\
\hline
\multirow{2}{*}{~~~$k=0$~~~} & $~~~B_u~~~$ & $6$ & $3$ & 0 & 0 & 
$1$ & 0 & 0\\
& $B_s$ & 0 &0 & $6$ & $3$ & 0 & 0 & $1$\\
\hline
$k=3$ & $B_u$ & 0 & 0 & 0 & 0 & $1$ & 0 & 0\\
\hline
\multirow{2}{*}{~~~$k=8$~~~} & $B_u$ & $6$ & $-6$ & 0 &0 & 
$1$ & 0 & 0\\
& $B_s$ & 0 & 0 & $6$ & $-6$ & 0 & 0 & $-2$\\
\hline\hline
\end{tabular}
\end{center}
\end{table}

\section{Coefficients for chiral one-loop contributions in $B$ baryon}
\label{sec:coefficients_B_baryon}

In this appendix, we present the coefficients in \Eq{wavefunction_baryon}, 
\Eq{tadpole_baryon}, and \Eq{sunset_baryon} 
corresponding to the self-energy, tadpole, and sunset one-loop diagrams. These coefficients are 
summarized in \Tab{wavefunction_baryon}, \Tab{tadpole_baryon}, \Tab{sunset_T_baryon}, 
\Tab{sunset_S_baryon_1}, and \Tab{sunset_S_baryon_2}. 

\begin{table}
\caption{%
\label{tab:wavefunction_baryon}
Coefficients $w^{T(S)}_{\phi_{ab}}$, $w'^{S}_{\phi_{ab}}$, $\tilde{w}^{T(S)}_{\phi_{aa}\phi'_{bb}}$, 
and $\tilde{w}'^{S}_{\phi_{aa}\phi'_{bb}}$ in \Eq{wavefunction_baryon} in the isospin limit. 
$\Sigma$, $\Xi$, and $\Xi'$ baryons with different 3-components of the isospin 
have the same coefficients because of the isospin symmetry. 
$\phi_{us}$ and $\phi_{su}$ are distinguished by the appearance of their mass parameters 
$\delta^{(B)}_{us}$ and $-\delta^{(B)}_{us}$ in the propagator of internal baryons. 
}
\begin{center}
\begin{tabular}{c|cccccccc|ccc}
&\multicolumn{8}{c|}{$w_\phi$}&\multicolumn{3}{c}{$w_{\phi\phi'}$}\\
\cline{2-12}
&~~~$uu$~~~&~~~$us$~~~&~~~$su$~~~&~~~$ss$~~~&~~~$uj$~~~&~~~$ur$~~~&~~~$sj$~~~&~~~$sr$~~~&
~$\eta_u\eta_u$~&~$\eta_u\eta_s$~&~$\eta_s\eta_s$~\\
\hline
~~~$\Lambda_b$~~~& 3 & 0 & 0 & 0 & 6 & 3 & 0 & 0 & 0 & 0 & 0\\
\hline
$\Xi_b$ & 0 & $\frac{3}{2}$ & $\frac{3}{2}$ & 0 & 3 & $\frac{3}{2}$ & 3 & $\frac{3}{2}$ 
& $\frac{1}{2}$ & $-1$ & $\frac{1}{2}$\\
\hline
$\Sigma$ & 1 & 0 & 0 & 0 & 2 & 1 & 0 & 0 & $\frac{2}{3}$ & 0 &0\\
\hline
$\Xi'$ & 0 & $\frac{1}{2}$ & $\frac{1}{2}$ & 0 & 1 & $\frac{1}{2}$ & 1 & $\frac{1}{2}$ 
& $\frac{1}{6}$ & $\frac{1}{3}$ & $\frac{1}{6}$\\
\hline
$\Omega$ & 0 & 0 & 0 & 1 & 0 & 0 & 2 & 1 & 0 & 0 & $\frac{2}{3}$\\
\hline\hline
&\multicolumn{8}{c|}{$w'_\phi$}&\multicolumn{3}{c}{$w'_{\phi\phi'}$}\\
\hline
$\Sigma$ & -1 & 0 & 0 & 0 & 2 & 1 & 0 & 0 & 0 & 0 &0\\
\hline
$\Xi'$ & 0 & $-\frac{1}{2}$ & $-\frac{1}{2}$ & 0 & 1 & $\frac{1}{2}$ & 1 & $\frac{1}{2}$ 
& $\frac{1}{6}$ & $-\frac{1}{3}$ & $\frac{1}{6}$\\
\hline
$\Omega$ & 0 & 0 & 0 & -1 & 0 & 0 & 2 & 1 & 0 & 0 & 0\\
\hline\hline
\end{tabular}
\end{center}
\end{table}

\begin{table}
\caption{%
\label{tab:tadpole_baryon}
Coefficients $x^{T(S),k}_{\phi_{ab}}$ and $\bar{x}^{T(S),k}_{\phi_{ab}}$ in \Eq{tadpole_baryon} 
in the isospin limit. 
For $k=0,8$, we find 
$x^{\Lambda_b}_{\phi}=x^{\Sigma_b}_{\phi}$, 
$\bar{x}^{\Lambda_b}_{\phi}=\bar{x}^{\Sigma_b}_{\phi}$, 
$x^{\Xi_b}_{\phi}=x^{\Xi'_b}_{\phi}$, 
and $\bar{x}^{\Xi_b}_{\phi}=\bar{x}^{\Xi'_b}_{\phi}$, 
where $\Sigma$, $\Xi$, and $\Xi'$ baryons with different 3-components of the isospin 
have the same coefficients because of 
isospin symmetry. For $k=3$, we find 
$x^{\Sigma_b^-}_\phi=-x^{\Sigma_b^+}_\phi$ and 
$x^{\Xi_b^{-\frac{1}{2}}}_\phi=x^{\Xi'^{-\frac{1}{2}}_b}_\phi
=-x^{\Xi_b^{+\frac{1}{2}}}_\phi=-x^{\Xi'^{+\frac{1}{2}}_b}_\phi$, 
while $x^{\Lambda_b}_\phi=x^{\Sigma_b^0}_\phi=x^{\Omega_b}_\phi=0$ 
for possible Goldstone mesons. 
Coefficients $\bar{x}$ are all zeros for $k=3$. 
}
\begin{center}
\begin{tabular}{c|c|cccc|cccc}
\multicolumn{2}{c|}{} & \multicolumn{4}{c|}{$x^{T(S),k}_\phi$}
& \multicolumn{4}{c}{$\bar{x}^{T(S),k}_{\phi}$}\\
\cline{3-10}
\multicolumn{2}{c|}{}&~~~$uj$~~~&~~~$ur$~~~&~~~$sj$~~~&~~~$sr$~~~&
~~~$uj$~~~&~~~$ur$~~~&~~~$sj$~~~&~~~$sr$~~~\\
\hline
\multirow{3}{*}{~~~$k=0$~~~} & $\Lambda_b$ & $-2$ & $-1$ & 0 & 0
& $2$ & $1$ & 0 & 0 \\
& ~~~$\Xi$~~~ & $-1$ & $-\frac{1}{2}$ & $-1$ & $-\frac{1}{2}$ 
& $1$ & $\frac{1}{2}$ & $1$ & $\frac{1}{2}$ \\
& $\Omega$ & 0 & 0 & $-2$ & $-1$
& 0 & 0 & $2$ & $1$ \\
\hline
\multirow{2}{*}{$k=3$} & $\Xi^{+\frac{1}{2}}$ & $-1$ & $-\frac{1}{2}$ & 0 & 0
& 0 & 0 & 0 & 0 \\
& $\Sigma^+$ & $-2$ & $-1$ & 0 & 0
& 0 & 0 & 0 & 0 \\
\hline 
\multirow{3}{*}{$k=8$} & $\Lambda_b$ & $-2$ & $-1$ & 0 & 0
& $2$ & $-2$ & 0 & 0 \\
& $\Xi$ & $-1$ & $-\frac{1}{2}$ & $2$ & $1$ 
& $1$ & $-1$ & $1$ & $-1$ \\
& $\Omega$ & 0 & 0 & $4$ & $2$
& 0 & 0 & $2$ & $-2$ \\
\hline\hline 
\end{tabular}
\end{center}
\end{table}

\begin{table}
\caption{%
\label{tab:sunset_T_baryon}
Coefficients $y^{T,k}_{\phi_{ab}}$, $\bar{y}^{T,k}_{\phi_{ab}}$ 
and $\tilde{y}^{T,k}_{\phi_{aa}\phi'_{bb}}$ in 
\Eq{sunset_baryon} in the isospin limit. 
The unbarred coefficients of $\Xi_b^{-\frac{1}{2}}$ are identical and opposite 
to those of $\Xi_b^{\frac{1}{2}}$ for $k=0,8$ and $k=3$, respectively; 
the unbarred coefficients of $\Lambda_b$ are all zeros for $k=3$. 
The barred coefficients of $\Xi_b^{-\frac{1}{2}}$ are identical 
to those of $\Xi_b^{\frac{1}{2}}$ for $k=0,8$, while the barred coefficients are all zeros for $k=3$. 
$\phi_{us}$ and $\phi_{su}$ are distinguished by the appearance of their mass parameters 
$\delta_{us}$ and $-\delta_{us}$ in the propagator of internal baryons. 
}
\begin{center}
\begin{tabular}{c|c|cccccccc|ccc}
\multicolumn{2}{c|}{}&\multicolumn{8}{c|}{$y^{T,k}_\phi$}&\multicolumn{3}{c}{$\tilde{y}^{T,k}_{\phi\phi'}$}\\
\cline{3-13}
\multicolumn{2}{c|}{}&~~~$uu$~~~&~~~$us$~~~&~~~$su$~~~&~~~$ss$~~~&~~~$uj$~~~&~~~$ur$~~~&~~~$sj$~~~&~~~$sr$~~~&
~$\eta_u\eta_u$~&~$\eta_u\eta_s$~&~$\eta_s\eta_s$~\\
\hline
\multirow{2}{*}{~~~$k=0$~~~} & ~~~$\Lambda_b$~~~ & $1$ & 0 & 0 
& 0 & $1$ & $\frac{1}{2}$ & 0 & 0 & 
$\frac{2}{3}$ & 0 & 0\\
&$\Xi_b^{+\frac{1}{2}}$ & 0 & $\frac{1}{2}$ & $\frac{1}{2}$ & 0 & 
$\frac{1}{2}$ & $\frac{1}{4}$ & $\frac{1}{2}$ & $\frac{1}{4}$ 
& $\frac{1}{6}$ & $\frac{1}{3}$ & $\frac{1}{6}$\\
\hline
$k=3$ & $\Xi_b^{+\frac{1}{2}}$ & 0 & 0 & $\frac{1}{2}$ 
& 0 & 0 & 0 & $\frac{1}{2}$ & $\frac{1}{4}$ 
& $\frac{1}{12}$ & $\frac{1}{6}$ & $\frac{1}{12}$\\
\hline
\multirow{2}{*}{~~~$k=8$~~~} & ~~~$\Lambda_b$~~~ & 
$1$ & 0 & 0 & 0 & $1$ & $\frac{1}{2}$ & 0 & 0 & 
$\frac{2}{3}$ & 0 & 0\\
&$\Xi_b^{+\frac{1}{2}}$ & 0 & $-1$ & $\frac{1}{2}$ & 0 & 
$-1$ & $-\frac{1}{2}$ & $\frac{1}{2}$ & $\frac{1}{4}$ 
& $-\frac{1}{12}$ & $-\frac{1}{6}$ & $-\frac{1}{12}$\\
\hline\hline
\multicolumn{2}{c|}{}&\multicolumn{8}{c|}{$\bar{y}^{T,k}_{\phi}$}&
\multicolumn{3}{c}{}\\
\cline{1-10}
\multirow{2}{*}{~~~$k=0$~~~} & ~~~$\Lambda_b$~~~ & 0 & 0 & 0 
& 0 & $1$ & $\frac{1}{2}$ & 0 & 0 & 
 &  & \\
&$\Xi_b^{+\frac{1}{2}}$ & 0 & 0 & 0 
& 0 & 
$\frac{1}{2}$ & $\frac{1}{4}$ & $\frac{1}{2}$ & $\frac{1}{4}$ 
&  &  & \\
\cline{1-10}
\multirow{2}{*}{~~~$k=8$~~~} & ~~~$\Lambda_b$~~~ & 
0 & 0 & 0 & 0 & $1$ & $-1$ & 0 & 0 & 
 &  & \\
&$\Xi_b^{+\frac{1}{2}}$ & 0 & 0 & 0 & 0 & 
$\frac{1}{2}$ & $-\frac{1}{2}$ & $\frac{1}{2}$ & $-\frac{1}{2}$ 
&  &  & \\
\hline\hline
\end{tabular}
\end{center}
\end{table}

\begin{table}
\caption{%
\label{tab:sunset_S_baryon_1}
Coefficients $y^{S,k}_{\phi_{ab}}$, $\bar{y}^{S,k}_{\phi_{ab}}$, 
and $\tilde{y}^{S,k}_{\phi_{aa}\phi'_{bb}}$ 
in \Eq{sunset_baryon} in the isospin limit. 
The unbarred coefficients of $\Sigma_b^-$, $\Xi'^-_b$ are 
identical and opposite to those of $\Sigma_b^+$, $\Xi'^+_b$ for $k=0,8$ and $k=3$, respectively; 
the unbarred coefficients of $\Sigma_b^0$ and $\Omega_b$ are all zeros for $k=3$. 
In the case of barred coefficients, 
$\Sigma$ and $\Xi'$ with different $3$-componets of the isospin 
have the same coefficients with those for $k=0,8$, while the barred coefficients are all zeros for $k=3$. 
$\phi_{us}$ and $\phi_{su}$ are distinguished by the appearance of their mass parameters 
$\delta_{us}$ and $-\delta_{us}$ in the propagator of internal baryons. 
}
\begin{center}
\begin{tabular}{c|c|cccccccc|ccc}
\multicolumn{2}{c|}{}&\multicolumn{8}{c|}{$y^{S,k}_\phi$}&\multicolumn{3}{c}{$\tilde{y}^{S,k}_{\phi\phi'}$}\\
\cline{3-13}
\multicolumn{2}{c|}{}&~~~$uu$~~~&~~~$us$~~~&~~~$su$~~~&~~~$ss$~~~&~~~$uj$~~~&~~~$ur$~~~&~~~$sj$~~~&~~~$sr$~~~&
~$\eta_u\eta_u$~&~$\eta_u\eta_s$~&~$\eta_s\eta_s$~\\
\hline
\multirow{4}{*}{~~~$k=0$~~~} 
& ~~~$\Sigma^0_b$~~~ & $1$ & 0 & 0 & 0 & 
$1$ & $\frac{1}{2}$ & 0 & 0 & $\frac{2}{3}$ & 0 & 0\\
& ~~~$\Sigma^+_b$~~~ & $1$ & 0 & 0 & 0 & 
$1$ & $\frac{1}{2}$ & 0 & 0 & 0 & 0 & 0\\
&$\Xi'^{+\frac{1}{2}}_b$ & 0 & $\frac{1}{2}$ & $\frac{1}{2}$ & 0 & 
$\frac{1}{2}$ & $\frac{1}{4}$ & $\frac{1}{2}$ & $\frac{1}{4}$ 
& $\frac{1}{6}$ & $\frac{1}{3}$ & $\frac{1}{6}$\\
& ~~~$\Omega_b$~~~ & 0 & 0 & 0 & $1$ & 
0 & 0 & $1$ & $\frac{1}{2}$ & 0 & 0 & 0\\
\hline
\multirow{2}{*}{$k=3$} 
& ~~~$\Sigma^+_b$~~~ & $1$ & 0 & 0 
& 0 & $1$ & $\frac{1}{2}$ & 0 & 0 & 0 & 0 & 0\\
& $\Xi'^{+\frac{1}{2}}_b$ & 0 & 0 & $\frac{1}{2}$ 
& 0 & 0 & 0 & $\frac{1}{2}$ & $\frac{1}{4}$ 
& $\frac{1}{12}$ & $\frac{1}{6}$ & $\frac{1}{12}$\\
\hline
\multirow{4}{*}{~~~$k=8$~~~} 
& ~~~$\Sigma^0_b$~~~ & $1$ & 0 & 0 & 0 & 
$1$ & $\frac{1}{2}$ & 0 & 0 & $\frac{2}{3}$ & 0 & 0\\
& ~~~$\Sigma^+_b$~~~ & $1$ & 0 & 0 & 0 & 
$1$ & $\frac{1}{2}$ & 0 & 0 & 0 & 0 & 0\\
&$\Xi'^{+\frac{1}{2}}_b$ & 0 & $-1$ & $\frac{1}{2}$ & 0 & 
$-1$ & $-\frac{1}{2}$ & $\frac{1}{2}$ & $\frac{1}{4}$ 
& $-\frac{1}{12}$ & $-\frac{1}{6}$ & $-\frac{1}{12}$\\
& ~~~$\Omega_b$~~~ & 0 & 0 & 0 & $-2$ & 
0 & 0 & $-2$ & $-1$ & 0 & 0 & 0\\
\hline\hline
\multicolumn{2}{c|}{}&\multicolumn{8}{c|}{$\bar{y}^{S,k}_{\phi}$}
&\multicolumn{3}{c}{}\\
\cline{1-10}
\multirow{3}{*}{~~~$k=0$~~~} 
& ~~~$\Sigma_b$~~~ & 0 & 0 & 0 & 0 & 
$1$ & $\frac{1}{2}$ & 0 & 0 & 
 &  & \\
&$\Xi'_b$ & 0 & 0 & 0 & 0 & 
$\frac{1}{2}$ & $\frac{1}{4}$ & $\frac{1}{2}$ & $\frac{1}{4}$ 
&  &  & \\
& ~~~$\Omega_b$~~~ & 0 & 0 & 0 & 0 & 
0 & 0 & $1$ & $\frac{1}{2}$ 
&  &  & \\
\cline{1-10}
\multirow{3}{*}{~~~$k=8$~~~} 
& ~~~$\Sigma_b$~~~ & 0 & 0 & 0 & 0 & 
$1$ & $-1$ & 0 & 0 & 
 &  & \\
&$\Xi'_b$ & 0 & 0 & 0 & 0 & 
$\frac{1}{2}$ & $-\frac{1}{2}$ & $\frac{1}{2}$ & $-\frac{1}{2}$ 
&  &  & \\
& ~~~$\Omega_b$~~~ & 0 & 0 & 0 & 0 & 
0 & 0 & $1$ & $-1$ & 
 &  & \\
\hline\hline
\end{tabular}
\end{center}
\end{table}

\begin{table}
\caption{%
\label{tab:sunset_S_baryon_2}
Coefficients $y'^{S,k}_{\phi_{ab}}$, $\bar{y}'^{S,k}_{\phi_{ab}}$, 
and $\tilde{y'}^{S,k}_{\phi_{aa}\phi'_{bb}}$ 
in \Eq{sunset_baryon} in the isospin limit. 
The unbarred coefficients of $\Sigma_b^-$, $\Xi'^-_b$ are 
identical and opposite to those of $\Sigma_b^+$, $\Xi'^+_b$ for $k=0,8$ and $k=3$, respectively; 
the unbarred coefficients of $\Sigma_b^0$ and $\Omega_b$ are all zeros for $k=3$. 
In the case of barred coefficients, 
$\Sigma$ and $\Xi'$ with different $3$-componets of the isospin 
have the same coefficients with those for $k=0,8$, while the barred coefficients are all zeros for $k=3$. 
$\phi_{us}$ and $\phi_{su}$ are distinguished by the appearance of their mass parameters 
$\delta_{us}$ and $-\delta_{us}$ in the propagator of internal baryons. 
}
\begin{center}
\begin{tabular}{c|c|cccccccc|ccc}
\multicolumn{2}{c|}{}&\multicolumn{8}{c|}{$y'^{S,k}_\phi$}&\multicolumn{3}{c}{$\tilde{y'}^{S,k}_{\phi\phi'}$}\\
\cline{3-13}
\multicolumn{2}{c|}{}&~~~$uu$~~~&~~~$us$~~~&~~~$su$~~~&~~~$ss$~~~&~~~$uj$~~~&~~~$ur$~~~&~~~$sj$~~~&~~~$sr$~~~&
~$\eta_u\eta_u$~&~$\eta_u\eta_s$~&~$\eta_s\eta_s$~\\
\hline
\multirow{4}{*}{~~~$k=0$~~~} 
& ~~~$\Sigma^0_b$~~~ & $1$ & 0 & 0 & 0 & 
$-1$ & $-\frac{1}{2}$ & 0 & 0 & $-\frac{2}{3}$ & 0 & 0\\
& ~~~$\Sigma^+_b$~~~ & $1$ & 0 & 0 & 0 & 
$-1$ & $-\frac{1}{2}$ & 0 & 0 & 0 & 0 & 0\\
&$\Xi'^{+\frac{1}{2}}_b$ & 0 & $\frac{1}{2}$ & $\frac{1}{2}$ & 0 & 
$-\frac{1}{2}$ & $-\frac{1}{4}$ & $-\frac{1}{2}$ & $-\frac{1}{4}$ 
& $-\frac{1}{6}$ & $-\frac{1}{3}$ & $-\frac{1}{6}$\\
& ~~~$\Omega_b$~~~ & 0 & 0 & 0 & $1$ & 0 
& 0 & $-1$ & $-\frac{1}{2}$ & 0 & 0 & 0\\
\hline
\multirow{2}{*}{$k=3$} 
& ~~~$\Sigma^+_b$~~~ & $1$ & 0 & 0 & 0 & 
$-1$ & $-\frac{1}{2}$ & 0 & 0 & 0 & 0 & 0\\
& $\Xi'^{+\frac{1}{2}}_b$ & 0 & 0 & $\frac{1}{2}$ & 0 & 
0 & 0 & $-\frac{1}{2}$ & $-\frac{1}{4}$ 
& $-\frac{1}{12}$ & $-\frac{1}{6}$ & $-\frac{1}{12}$\\
\hline
\multirow{4}{*}{~~~$k=8$~~~} 
& ~~~$\Sigma^0_b$~~~ & $1$ & 0 & 0 & 0 & 
$-1$ & $-\frac{1}{2}$ & 0 & 0 & $-\frac{2}{3}$ & 0 & 0\\
& ~~~$\Sigma^+_b$~~~ & $1$ & 0 & 0 & 0 & 
$-1$ & $-\frac{1}{2}$ & 0 & 0 & 0 & 0 & 0\\
&$\Xi'^{+\frac{1}{2}}_b$ & 0 & $-1$ & $\frac{1}{2}$ & 0 & 
$1$ & $\frac{1}{2}$ & $-\frac{1}{2}$ & $-\frac{1}{4}$ 
& $\frac{1}{12}$ & $\frac{1}{6}$ & $\frac{1}{12}$\\
& ~~~$\Omega_b$~~~ & 0 & 0 & 0 & $-2$ & 0 
& 0 & $2$ & $1$ & 0 & 0 & 0\\
\hline\hline
\multicolumn{2}{c|}{}&\multicolumn{8}{c|}{$\bar{y}'^{S,k}_{\phi}$}
&\multicolumn{3}{c}{}\\
\cline{1-10}
\multirow{3}{*}{~~~$k=0$~~~} 
& ~~~$\Sigma_b$~~~ & 0 & 0 & 0 & 0 
& $-1$ & $-\frac{1}{2}$ & 0 & 0 & 
 &  & \\
&$\Xi'_b$ & 0 & 0 & 0 & 0 & 
$-\frac{1}{2}$ & $-\frac{1}{4}$ & $-\frac{1}{2}$ & $-\frac{1}{4}$ 
&  &  & \\
& ~~~$\Omega_b$~~~ & 0 & 0 & 0 & 0 & 
0 & 0 & $-1$ & $-\frac{1}{2}$ & 
 &  & \\
\cline{1-10}
\multirow{3}{*}{~~~$k=8$~~~} 
& ~~~$\Sigma_b$~~~ & 0 & 0 & 0 & 0 & 
$-1$ & $1$ & 0 & 0 & 
 &  & \\
&$\Xi'_b$ & 0 & 0 & 0 & 0 & 
$-\frac{1}{2}$ & $\frac{1}{2}$ & $-\frac{1}{2}$ & $\frac{1}{2}$ 
&  &  & \\
& ~~~$\Omega_b$~~~ & 0 & 0 & 0 & 0 & 
0 & 0 & $-1$ & $1$ & 
 &  & \\
\hline\hline
\end{tabular}
\end{center}
\end{table}

\end{document}